\documentclass{article}

\usepackage{graphicx} 

\usepackage{natbib}

\usepackage{algorithm}
\usepackage{algorithmic}

\usepackage{amsmath, amsfonts, latexsym, amsthm, amssymb}
\usepackage{multirow}
\usepackage{pdfsync}
\usepackage{float}
\usepackage[caption=false]{subfig}
\usepackage{amsmath}
\usepackage{amssymb}
\usepackage{epsfig}

\newcounter{xxx}
\newcommand\observations{{\mathcal Y}}
\newcommand\states{{\mathcal X}}

\def\N{{\mathbb N}}        

\def\P{{\mathbb P}}        
\def\1{{\mathbf 1}}        

\newtheorem{theorem}{Theorem}

\newtheorem{proposition}[theorem]{Proposition}

\newcommand{\ud}{\,\mathrm{d}}

\usepackage{url}

\usepackage{hyperref}

\usepackage{booktabs,fancyhdr} 

\usepackage[accepted]{icml2014}

\icmltitlerunning{Efficient Continuous-Time Markov Chain Estimation}

\begin{document} 

\twocolumn[
\icmltitle{Efficient Continuous-Time Markov Chain Estimation}

{\footnotesize
\icmlauthor{Monir Hajiaghayi}{monirh@cs.ubc.ca}
\icmladdress{\footnotesize Department of Computer Science, University of British Columbia, Vancouver, BC V6T 1Z4, Canada}
\icmlauthor{Bonnie Kirkpatrick}{bbkirk@cs.miami.edu}
\icmladdress{\footnotesize Department of Computer Science, University of Miami, Coral Gables, FL 33124, United States}
\icmlauthor{Liangliang Wang }{liangliang\_wang@sfu.ca}
\icmladdress{\footnotesize Department of Statistical and Actuarial Sciences, Simon Fraser University, Burnaby, BC V5A 1S6, Canada}
\icmlauthor{Alexandre Bouchard-C\^ot\'e }{bouchard@stat.ubc.ca}
\icmladdress{\footnotesize Statistics Department, University of British Columbia, Vancouver, BC V6T 1Z4, Canada}}
\icmlkeywords{CTMCs, combinatorial space, applications}
 
\vskip 0.25in
]

\begin{abstract} 
Many problems of practical interest rely on Continuous-time Markov chains~(CTMCs) defined over combinatorial state spaces, rendering the computation of transition probabilities, and hence probabilistic inference, difficult or impossible with existing methods. 
For problems with countably infinite states, where classical methods such as matrix exponentiation are not applicable, 
the main alternative has been particle Markov chain Monte Carlo methods imputing both the holding times and sequences of visited states.   
We propose a particle-based Monte Carlo approach where the holding times are marginalized analytically. 
We demonstrate that in a range of realistic inferential setups, our scheme dramatically reduces the variance of the Monte Carlo approximation and yields more accurate parameter posterior approximations given a fixed computational budget.
These experiments are performed on both synthetic and real datasets, drawing from two important examples of CTMCs having combinatorial state spaces: string-valued mutation models in phylogenetics and nucleic acid folding pathways. 
\end{abstract} 
\vspace{-0.4cm}
\section{Introduction}

Continuous-time Markov chains (CTMCs) play a central role in applications as diverse as queueing theory,
phylogenetics, genetics, and models of chemical interactions~\cite{huelsenbeck_mrbayes:_2001,Munsky2006}. 
The process can be thought of as a timed random walk on a directed graph where the countable, but potentially infinite, set of graph nodes are the values that the process can take on. There are probabilities of transition associated with the edges of the graph, and the holding time, or length of time between two transitions, is exponentially distributed with a rate depending on the current node.  A path simulated from this random process is an ordered list of the nodes visited and the times at which they are reached.  


In leveraging the modelling capabilities of CTMCs, the bottleneck is typically the computation of the \emph{transition probabilities}: the conditional probability that a trajectory ends in a given end state, given a start state and a time interval. This computation involves the marginalization over the uncountable set of end-point conditioned paths. Although we focus on the Bayesian framework in this work, where the transition probabilities appear in Metropolis-Hastings ratios, the same bottleneck is present in the frequentist framework, where transition probabilities are required for likelihood evaluation. 
When the state space is small, exact marginalization can be done analytically 
 via the matrix exponential. 
Unfortunately, this approach is not directly applicable to infinite state spaces, and is not computationally feasible in large state spaces because of the cubic running time of matrix exponentiation.

We propose an efficient Monte Carlo method to approach inference in CTMCs with weak assumptions on the state space. Our method can approximate transition probabilities as well as estimate CTMC parameters for this general class of processes. 
More precisely, we are interested in countably infinite state space CTMCs that satisfy the following two criteria. First, we require the construction of a certain type of potential on the state space. We describe this potential in more detail in Section~\ref{section:Method}, and show in Section~\ref{sec:numerical-examples} that such potentials can be easily constructed even for complex models. Second, the CTMC should be explosion-free to avoid pathologies (i.e., we require that there is a finite number of transitions with probability one in any bounded time interval). 

In contrast, classical uniformization methods assume that there is a fixed bound on all the rates \cite{Grassmann1977Uniformization},  
a much stronger condition than our explosion-free assumption. For example, in the first of the two application domains that we investigated, inference in string-valued CTMCs for phylogenetics,  the models are explosion-free but do not have a fixed bound on the rates. 
Other approaches, based on performing Markov chain Monte Carlo~(MCMC) with auxiliary variables, relax the bounded rate assumption \cite{Rao2011,Rao2012}, but they have a running time that depends linearly on the size of the state space in the sparse case and quadratically in the dense case. 

Particle-based methods offer an interesting complementary approach, as they have a time complexity per particle that depends on the imputed \emph{number of transitions between the two end points} instead of on the size of the state space. 

In the simplest case, one can implement this idea using a proposal distribution equal to the generative process over paths initialized at the start point. The weight of a particle is then equal to one if the end point of the generated path coincides with observed end point, and zero otherwise. We call this proposal the \emph{forward sampling} proposal. This idea can be turned into a consistent estimator of posterior distributions over parameters using pseudo-marginal methods \cite{Beaumont2003GIMH,Andrieu2009PseudoMarg} (or in more complicated setups, particle MCMC methods \cite{Andrieu2010}).


Unfortunately, the forward sampling method has two serious limitations. First, the requirement of imputing waiting times between each transition means that the proposal distribution is defined over a potentially high-dimensional continuous space. This implies that large numbers of particles are required in practice. Second, in problems where each state has a large number of successors, the probability of reaching the end state can  become extremely small, which for example further inflates the number of particles required to obtain non degenerate Metropolis-Hastings ratios in particle MCMC~\cite{Andrieu2010} algorithms. 

End point informed proposals over transitions and waiting times have been developed in previous work \cite{fan2008sampling}, but  this previous work is tailored to dynamic Bayesian models rather than to the combinatorial problems studied here. Our method greatly simplifies the development of end point informed proposals  by marginalizing all continuous variables.
There has also been work on related end-point conditioning problems in the rare event simulation literature \cite{Juneja2006RareEvent}, but this previous work has focused on the discrete-time setting.

\section{Methodology}
\label{section:Method}

For expositional purposes, we start by describing the simplest setup in which our method can be applied: computing the probability that a CTMC with known rate parameters occupies state $y\in\states$ at time $T$ given that it occupies state $x\in\states$ at time 0, where $\states$ is a countable set of states. The main contributions of this paper can be understood in this simple setup. We then show that our method can be extended to certain types of partial or noisy observations, to more than two observations organized as a time series or a tree (branching process), and to situations where some or all the parameters of the CTMC are unknown.   

{\bf Notation.} Let $\nu(x,y)$ denote the transition probability from state $x\in\states$ to state $y\in\states$ given that a state jump occurs (i.e. $\sum_{y:y\neq x} \nu(x,y) = 1, \nu(x,x) = 0$). Let $\lambda(x)$ denote the rate of the exponentially-distributed holding time at state $x$ ($\lambda : \states \to [0, \infty)$).\footnote{Note that this is a reparameterization of the standard rate matrix $q_{x,y}$, with $q_{x,x} = -\lambda(x)$, and $q_{x,y} = \lambda(x) \nu(x,y)$ for $x\neq y$.} We only require efficient point-wise evaluation of $\lambda(\cdot),\nu(\cdot,\cdot)$ and efficient simulation from $\nu(x,\cdot)$ for all $x\in\states$. We start by assuming that $\nu$ and $\lambda$ are fixed, and discuss their estimation afterward. 
We define some notation for paths sampled from this process. Let $X_1, X_2, \dots$ denote the list of visited states with $X_i \neq X_{i+1}$, called the \emph{jump chain}, and $H_1, H_2, \dots$, the list of corresponding \emph{holding times}. The model is characterized by the following distributions: $X_{i+1}|X_{i} \sim \nu(X_i, \cdot)$, $H_i|X_i \sim F(\lambda(X_i))$, where $F(\lambda)$ is the exponential distribution CDF with rate $\lambda$. Given a start state $X_1=x$, we denote by $\P_x$ the probability distribution induced by this model. Finally, we denote by $N$ the number of states visited, counting multiplicities, in the interval $[0,T]$, i.e. $(N = n) = (\sum_{i=1}^{n-1} H_i \le T < \sum_{i=1}^n H_i)$.



{\bf Overview of the inference method.} Using the simple setup introduced above, the problem we try to solve is to approximate $\P_x(X_N = y)$, which we approach using an importance sampling method.  Each proposed particle consists of a sequence (a list of variable finite length) of states, $x^*=(x_1, \dots, x_n)\in\states^*$, starting at $x$ and ending at $y$. In other words, we marginalize the holding times, hence avoiding the difficulties involved with sequentially proposing times constrained to sum to the time $T$ between the end points.

Concretely, our method is based on the following elementary property, proved in the Supplement:
\begin{proposition}\label{prop:first}
If we let $\pi(x^*) = \gamma(x^*)/\P_x(X_N = y)$, where, {\footnotesize \begin{flalign}\label{eq:main}
\gamma(x^*) = &\1(x_n = y) \left(\prod_{i=1}^{n-1} \nu(x_i, x_{i+1})\right) \times \\
& \P\left(\sum_{i=1}^{n-1} H_i \le T < \sum_{i=1}^n H_i\bigg|X^* = x^*\right), \nonumber
\end{flalign}} where the $H_i$'s are sampled according to $F(\lambda(X_i))$ independently given $X^*= (X_1, \cdots,X_N)$ and where $n=|x^*|$, 
then $\pi$ is a normalized probability mass function.
\end{proposition}

%

As our notation for $\gamma,\pi$ suggests, we use this result as follows (see Algorithm 1 in the Supplement for details). First, we define an importance sampling algorithm that targets the unnormalized density $\gamma(x^*)$ via a proposal $\tilde\P(X^*=x^*)$. Let us denote the $k$-th particle produced by this algorithm by $x^*(k)\in \states^*$, $k\in \{1, \dots, K\}$, where the number of particles $K$ is an approximation accuracy parameter. Each of the $K$ particles is sampled independently according to the proposal $\tilde\P$.  Second, we exploit the fact that the sample average of the unnormalized importance weights $w(x^*(k)) = \gamma(x^*(k))/\tilde\P(X^*=x^*(k))$ generated by this algorithm provide a consistent estimator for the normalizer of $\gamma$. Finally, by Proposition~\ref{prop:first}, this normalizer coincides with the quantity of interest here, $\P_x(X_N = y)$.  The only formal requirement on the proposal is that $\P_x(X^*=x^*) >0$ should imply $\tilde\P(X^*=x^*) >0$. However, to render this algorithm practical, we need to show that it is possible to define efficient proposals, in particular proposals such that $\P_x(X^*=x) >0$ if and only if $\tilde\P(X^*=x^*) >0$ (in order to avoid particles of zero weight). We also need to show that $\gamma$ can be evaluated point-wise efficiently, which we establish in Proposition~\ref{prop:matrixq}.





%
%
%

%
%
%
%
%
%

{\bf Proposal distributions.}
Our proposal distribution is based on the idea of simulating from the jump chain, i.e.~of sequentially sampling from $\nu$ until $y$ is reached. However this idea needs to be modified for two reasons. First, (1) since the state is countably infinite in the general case, there is a potentially positive probability that the jump chain sampling procedure will never hit $y$. Even when the state is finite, it may take an unreasonably large number of steps to reach $y$. Second, (2) forward jump chain sampling, assigns zero probability to paths visiting $y$ more than once. 

We address (1) by using a user-specified \emph{potential} $\rho^y : \states \to \N$ 
centred at the target state $y$ (see Supplement for the conditions we impose on $\rho^y$). 
For example we used the Levenshtein (i.e., minimum number of insertion, deletion, and substitution required to change one string into another) and Hamming distances for the string evolution and RNA kinetics applications respectively. 
Informally, the fact that this distance favors states which are closer to $y$ is all that we need to bias the sampling of our new jump process towards visiting $y$.


How do we bias the proposal sampling of the next state?  Let $D(x)\subset\states$ be the set of states that decrease the potential from $x$.  The proposed jump-chain transitions are chosen with probability  
{\footnotesize\begin{flalign}\label{eq:prop}
 &{\tilde \P(X_{i+1} = x_{i+1} | X_i = x_i) =} \\ & (\alpha_{x_i}^y) \left(\frac{ \nu(x_i,x_{i+1}) \1\{x_{i+1} \in D(x_i)\}}{ \sum_{x'_{i+1} \in D(x_i)} \nu(x_i,x'_{i+1})}\right) \nonumber\\ & +  (1-\alpha_{x_i}^y) \left( \frac{\nu(x_i,x_{i+1})(1-\1\{x_{i+1} \in D(x_i)\}) }{ \sum_{x'_{i+1} \notin D(x_i)} \nu(x_i,x'_{i+1})}\right) . \nonumber
\end{flalign}}
We show in the Supplement that under weak conditions, we will hit target $y$ in finite time with probability one if we pick $\alpha_{x}^y = \max \{ \alpha,  \sum_{x'_{i+1} \in D(x_i)} \nu(x_i,x'_{i+1})\}$. Here $\alpha > 1/2$ is a tuning parameter. We discuss the sensitivity of this parameter, as well as strategies for setting it in Section~\ref{sec:RFP}.



Point (2) can be easily addressed by simulating a geometrically-distributed number of excursions where the first excursion starts at $x$, and the others at $y$, and each excursion ends at $y$. We let $\beta$~denote the parameter of this geometric distribution, a tuning parameter, which we also discuss at the end of Section~\ref{sec:RFP}. 
%
%

\newcommand\xspace{{\mathcal X}}

{\bf Analytic jump integration.} In this section, we describe how the unnormalized density $\gamma(x^*)$ defined in Equation~(\ref{eq:main}) can be evaluated efficiently for any given path $x^* \in \xspace^*$. 

It is enough to show that we can compute the following integral for $H_i|X^*\sim F(\lambda(X_i))$ independently conditionally on $X^*$: {\footnotesize \begin{flalign}\label{eq:matrixq}
&\P\left(\sum_{i=1}^{n-1} H_i \le T < \sum_{i=1}^n H_i\bigg|X^* = x^*\right) =\\ 
 &\;\;\;\idotsint_{h_i > 0 : \sum_{i=1}^n h_i = T} g(h_1, h_2, \dots, h_{n}) \ud h_1 \ud h_2 \dots \ud h_{n}, \nonumber\\
&\textrm{where}\;\;\;\;\;g(h_1, h_2, \dots, h_{n}) =\nonumber\\ & \;\;\;\left\{\prod_{i=1}^{n-1} f(h_i; \lambda(x_i)) \right\} (1-F(h_{n} ; \lambda(x_n))),\nonumber
\end{flalign}}
 and where $f$ is the exponential density function.
Unfortunately, there is no efficient closed form for this high-dimensional integral, except for special cases (for example, if all rates are equal)~\cite{Akkouchi2008}. This integral is related to those needed for computing convolutions of non-identical independent  exponential random variables.
While there exists a rich literature on numerical approximations to these convolutions, 
these methods either add assumptions on the rate multiplicities (e.g. $|\{\lambda(x_1), \dots, \lambda(x_N)\} |= |(\lambda(x_1), \dots, \lambda(x_N))|$), or are computationally intractable \cite{Amari1997SumExp}. 

We propose to do this integration using the construction of an auxiliary, finite state CTMC with a $n+1$ by $n+1$ rate matrix $\check Q$ (to be defined shortly). The states of $\check Q$ correspond to the states visited in the path $(x_1, x_2, \dots, x_n)$ with multiplicities plus an extra state $s_{n+1}$. All off-diagonal entries of $\check Q$ are set to zero with the exception of transitions going from $x_i$ to $x_{i+1}$, for $i\in\{1, \dots, n\}$.  More specifically,  $\check{Q}$ is 
{\footnotesize \begin{eqnarray}\label{eq:tildeq}
\left[ \begin{array}{llllll} 
-\lambda(x_1)  &  \lambda(x_1)  & 0  &\cdots & 0  & 0\\
0& -\lambda(x_2)  &  \lambda(x_2)  &\cdots & 0  & 0  \\
\cdots & \cdots& \cdots  & \cdots  & \cdots   & \cdots \\
  0 & 0& 0&\cdots& -\lambda(x_{n})   &  \lambda(x_{n})     \\
    0 & 0& 0&\cdots& 0  &  0    \\
 \end{array} \right].
\end{eqnarray}}

This construction is motivated by the following property  which is proven in the Supplement:
\begin{proposition} 
\label{prop:matrixq}
For any finite proposed path $(x_1, x_2, \dots, x_n)$, if $\check Q$ is defined as in Equation~(\ref{eq:tildeq}), then
{\footnotesize
\begin{flalign}
&\left(\exp(T \check Q)\right)_{1,n} = \P\left(\sum_{i=1}^{n-1} H_i \le T < \sum_{i=1}^n H_i\bigg|X^* = x^*\right)
\end{flalign}}where $\exp(A)$ denotes the matrix exponential of $A$.\footnote{Multiplicities of the rates in $\check Q$ greater than one will break diagonalization-based methods of solving $\exp(T \check Q)$, but other efficient matrix exponentiation methods such as the squaring and scaling method are still applicable in these cases.}
\end{proposition}





{\bf Trees and sequences of observation.} We have assumed so far that the observations take the form of a single branch with the state fully observed at each end point. 
To approach more general types of observations, for example a series of partially observed states, or a phylogenetic tree with observed leaves, our method can be generalized by replacing the importance sampling algorithm by a sequential Monte Carlo (SMC) algorithm. 
We focus on the tree case in this work which we describe in detail in Section~\ref{section:Numerical examples}, but we outline here how certain partially observed sequences can also be approached to start with something simpler.

Consider a setup where the observation at time $T_i$ is a set $A_i \subset \states$ (i.e. we condition on $(X(T_i) \in A_i; i\in\{1,\dots,m\})$ which arises for example in \cite{Saeedi2011Priors}). In this case, the importance sampling algorithm described in the previous section  can be used at each iteration, with the main difference being that the potential $\rho$ is modified to compute a  distance to a set $A_i$ rather than a distance to a single point $y$. See Algorithm 5 in the Supplement for details.

For other setups, the construction of the potential is more problem-specific. One limitation of our method arises when the observations are only weakly informative of the hidden state. We leave these difficult instances for future work and reiterate that many interesting and challenging problems fall within the reach of our method (for example, the computational biology problems presented in the next section).




%
%
%

{\bf Parameter estimation.} So far, we have assumed that the parameters $\nu$ and $\lambda$ governing the dynamics of the process are known. We now consider the case where we have a parametric family with unknown parameter $\theta \in \Theta$ for the jump transition probabilities $\nu_\theta$ and for the holding time mean function $\lambda_\theta$. We denote by $\P_{x,\theta}$ the induced distribution on paths and by $p$ a prior density on $\theta$. To approximate the posterior distribution on $\theta$, we use pseudo-marginal methods \cite{Beaumont2003GIMH,Andrieu2009PseudoMarg} in the fixed end-point setup and particle MCMC methods \cite{Andrieu2010} in the sequences and trees setup. While our algorithm can be combined with many variants of these pseudo-marginal and particle MCMC methods, in this section,  for simplicity we describe the grouped independence Metropolis-Hastings (GIMH) approach.


At each MCMC iteration $t$, the algorithm keeps in memory a pair $x^{(t)} = (\theta^{(t)}, \hat Z^{(t)}_{\theta^{(t)}})$ containing a current parameter $\theta^{(t)}$ and an approximation $\hat Z^{(t)}_{\theta^{(t)}}$ of the marginal probability of the observations\footnote{For example, in the single branch setting, $\observations = (X_1 = x, X_N = y)$.} $\observations$ given $\theta^{(t)}$, $\hat Z^{(t)}_{\theta^{(t)}} \approx \P_{\theta^{(t)}}(\observations)$. This approximation is obtained from the algorithm described in the previous subsections. Even though this approximation is inexact for a finite number of particles, the GIMH sampler is still guaranteed to converge to the correct stationary distribution~\cite{Andrieu2010}. 

The algorithm requires the specification of a proposal density on parameter $q(\theta'|\theta)$. At the beginning of each MCMC iteration, we start by proposing a parameter $\theta^*$ from this proposal $q$. We then use the estimate $\hat Z_{\theta^*}$ of $\P_{\theta^*}(\observations)$ given by the average of the weights $w(x^{*(t)}(k))$ to form the ratio $r(\theta^{(t)}, \theta^*)$, below, where $k$ is the index for particles.  We accept $(\theta^{*}, \hat Z_{\theta^{*}})$, or remain as before, according to a Bernoulli distribution with probability $\min\{1, r(\theta^{(t)}, \theta^*)\}$ where 
 \begin{align*} 
r(\theta^{(t)}, \theta^*) = \frac{p(\theta^*)}{p(\theta^{(t)})} \ \frac{\hat Z_{\theta^*}}{\hat Z^{(t)}_{\theta^{(t)}}} \ \frac{q(\theta^{(t)} | \theta^*)}{q( \theta^* | \theta^{(t)})}.
\end{align*}See Algorithm 4 in the Supplement for details.


%




\section{Numerical examples}
\label{sec:numerical-examples}
\label{section:Numerical examples}

 \newcommand{\seqstate}{\boldsymbol{\Im}}
 \newcommand{\str}{x}
\newcommand{\eventsorder}{O}
\newcommand{\ssetypes}{\mathcal{T}}
\newcommand{\muts}{M}
\newcommand{\setmutations}{\mathcal{M}}
\newcommand{\node}{v}
\newcommand{\subtree}{f}
\newcommand{\hidden}{X}
\newcommand{\pstate}{s}

\subsection{String-valued evolutionary models}



Molecular evolutionary models, central ingredients of modern phylogenetics, describe how biomolecular sequences (RNA, DNA, or proteins) evolve over time via a CTMC where jumps are character substitutions, insertions and deletions (indel), and states are biomolecular sequences. 
Previous work focused on the relatively restricted
range of evolutionary phenomena for which computing  marginal probabilities of the form $\P_x(X_N = y)$ can be done exactly.

In particular, we are not aware of existing methods for doing Bayesian inference over  context-dependent indel models, i.e. models where insertions and deletions can depend on flanking characters. Modelling the context of indels is important because of a phenomenon called  \emph{slipped strand mispairing} (SSM), a well known explanation for the evolution of repeated sequences \cite{Morrison2009,Hickey2011b,Arribas-Gil2012}. For example, if  a  DNA string contains a substring of  \textit{``TATATA"}, the non-uniform error distribution in DNA replication is likely to lead to a long insertion of extra \textit{``TA''} repeats.

\begin{figure*}[tp]
\begin{center}
  \includegraphics[width=6.5in]{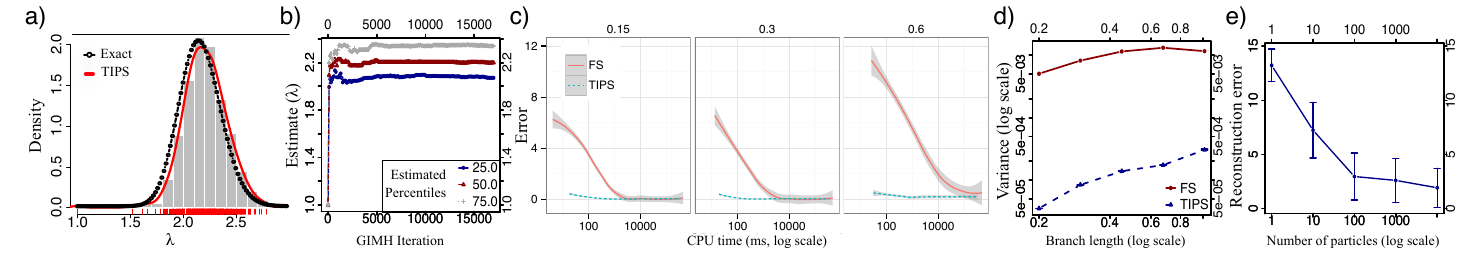}
  \caption{{\footnotesize a) Validation of the posterior estimate on the Poisson Indel Process dataset. The histogram and the density estimate in red is obtained from 35,000 GIMH iterations; the black curve is obtained by numerical integration. The generating value is $\lambda_\textrm{pt} = 2$.  b) Convergence of percentiles computed from prefixes of the GIMH output. c) Relative errors on the transition probabilities for branch lengths from $\{0.15,0.3,0.6\}$. d) Estimated variance of the weights. e) Reconstruction error on tree distances.  
}}
  \label{fig:phylo}
\end{center} 
\end{figure*}

{\bf Model.} In order to describe our SSM-aware model, it is enough to describe its behavior on a single branch of a tree, say of length $T$.
Each marginal variable $X_t$ is assumed
to have the countably infinite domain of all possible molecular sequences.
We define 
 $\lambda(x)$, as a function of the mutation rate per base
 $\theta_\textrm{sub}$, the global point insertion (i.e. insertion of a single nucleotide) rate $\lambda_\textrm{pt}$, the point deletion rate per base $\mu_\textrm{pt}$, the global
SSM  insertion rate $\lambda_\textrm{SSM}$ (which copies a substring of length up to three to the right of that substring), and the SSM deletion rate per
 valid  SSM deletion location $\mu_\textrm{SSM}$ (deletion of a substring of length up to three at the right of an identical substring):
\begin{align}
 \lambda(x)= m(x)\theta_\textrm{sub} +  \lambda_\textrm{pt} + m(x)\mu_\textrm{pt} + \lambda_\textrm{SSM} +
k(x)\mu_\textrm{SSM} \nonumber
\end{align} where $m(x)$ is the length of the string $x$ and $k(x)$ is the number of valid SSM
deletion locations in $x$. We denote these evolutionary parameters by $\theta=(\theta_\textrm{sub}, \lambda_\textrm{pt}, \mu_\textrm{pt}, \lambda_\textrm{SSM}, \mu_\textrm{SSM})$. 
The jump transition probabilities from $x$ to $x'$ are obtained by normalizing each of the above rates. For example the probability of deleting the first character given that there is a change from sequence $x$ is $\mu_\textrm{pt} / \lambda(x)$. 
Note that since the total insertion rate does not depend on the length of the string, the process is explosion-free for all $\theta$. At the same time, there is no fixed bound on the deletion rate, ruling out classical methods such as uniformization or matrix exponentiation.

{\bf Validation on a special case.} Before moving on to more complex experiments, we started with a special case of our model where the true posterior can be computed numerically. This is possible by picking a single branch, and setting $\lambda_\textrm{SSM} = \mu_\textrm{SSM} = 0$, in which case the process reduces to a process for which analytic calculation of $\P_{x,\theta}(X_N = y)$ is tractable \cite{Bouchard2012Evolutionary}. We fixed the substitution parameter $\theta_\textrm{sub}$, and computed as a reference the posterior by numerical integration
on $\lambda_\textrm{pt}$, $\mu_\textrm{pt}$ truncated to $[0, 3]^2$ and using $100^2$ bins.

We generated 200 pairs of sequences along with their sequence alignments\footnote{A sequence alignment is a graph over the observed nucleotides linking the nucleotides that have a common ancestor.}, with $T = 3/10, \lambda = \lambda_\textrm{pt} = 2, \mu = \mu_\textrm{pt} = 1/2$ and held out the mutations and the true value of parameters $\lambda$ and $\mu$. We put an exponential prior with rate 1.0 on each parameter.  We approximate the posterior using our method, initializing the parameters to $\lambda = \mu = 1$, using $\alpha = 2/3, \beta = 19/20$, $64$ particles, and a proposal $q$ over parameters given by the multiplicative proposal of Lakner et al.~\cite{Lakner01022008}.  We show the results of $\lambda$ in Figure~\ref{fig:phylo}a 
 and in the Supplement Figure: results of parameter $\mu$. In both cases the posterior approximation is shown to closely mirror the numerical approximation. The evolution of the Monte Carlo quartiles computed on the prefixes of Monte Carlo samples also shows that the convergence is rapid (Figure~\ref{fig:phylo}b).

Next, we compared the performance of a GIMH algorithm 
computing $\hat Z_{\theta^{(t)}}$ using our method, with a GIMH algorithm computing $\hat Z^{(t)}_{\theta^{(t)}}$ using forward sampling. 
We performed this comparison by computing the Effective Sample Size (ESS) after a fixed computational budget (3 days). For the parameter $\lambda$, our method achieves an ESS of 1782.7 versus 44.6 for the forward sampling GIMH method; for the parameter $\mu$, our method achieves an ESS of 6761.2 versus 90.2 for the forward sampling GIMH method. In those experiments, we used 100 particles per MCMC step, but we also tried different values and observed the same large gap favoring our method (see Supplement Figure: Varying the number of particles per MCMC step).

%
%

%

We also generated three datasets based on branch lengths from $\{0.15,0.3,0.6\}$, each containing 10 pairs of sequences $(x,y)$ along with their sequence alignments and estimated the transition probability $\P_x(Y_N = y)$ using our method (denoted Time-Integrated Path Sampling, TIPS), and using forward simulation (denoted forward sampling, FS). We compared the two methods in Figure~\ref{fig:phylo}c by looking at the absolute log error of the estimate $\hat p$, error($\hat p$)~$=|\log \hat p - \log \P_x(X_N = y)|$. We performed this experiment with a range of numbers of particles, $\{2^1,2^2, \dots, 2^{20}\}$ and plotted the relative errors as a function of the wall clock time needed for each approximation method.
We also computed the variances of the importance weights for specific alignments and compared these variances for FS and TIPS (see Figure~\ref{fig:phylo}d). We observed that the variances were consistently two orders of magnitudes lower with our method compared to FS.


%

{\bf Tree inference via SMC.} We now consider the general case, where inference is on a phylogenetic tree, and the SSM parameters are non-zero.  
To do this, we use existing SMC algorithms for phylogenetic trees \cite{teh08a,Bouchard2012Phylogenetic,Wang2012PhDThesis}, calling our algorithm at each proposal step. We review phylogenetic inference in the Supplement where we also give in Algorithm 6 the details of how we combined our method with phylogenetic SMC.

To evaluate our method,  we sampled 10 random trees from the coalescent on 10 leaves,  along each of
which we simulated 5 sets of molecular sequences according to our evolutionary model. We used the following parameters:  SSM length=3, 
$\theta_{sub}=0.03$, $\lambda_{pt}=0.05$,  $\mu_{pt}=0.2$, 
 $\lambda_{SSM}=2.0$, and 
 $\mu_{SSM}=2.0$.  One subset of simulated
 data is shown in the Supplement Figure: Sequence Simulation. 
 The unaligned sequences on leaves are used for tree reconstruction using our method. 
 We summarized the
 posterior over trees using a consensus tree optimizing the posterior expected pairwise distances \cite{Felsenstein1981}. Figure~\ref{fig:phylo}e shows  tree distances using the partition metric \cite{Felsenstein2003}
 between 
 generated trees and consensus trees reconstructed using our evolutionary model.
 The tree distance decreases as the number of particles increases, and a reasonable accuracy is obtained with only 100 particles, 
 suggesting that it is
 possible to reconstruct phylogenies from noisy data 
 generated by complex evolutionary mechanisms.

\subsection{RNA folding pathways}
\label{sec:RFP}

Nucleic acid folding pathways predict how RNA and DNA molecules
fold in on themselves via intra-molecular interactions.  The state
space of our stochastic process that describes folding is the set of
all folds, or secondary structures, of the nucleic acid
molecule which is a combinatorial object.  
For RNA molecules, the secondary structure is the primary determiner of RNA function.  For DNA its fold can help determine gene transcription rates.
Understanding the folding pathways can be useful for  designing
nano-scale machines that have potential health applications \cite{Venkataraman2010}.
For these reasons, it is often useful in applications to get an accurate estimate of the probability that a nucleic acid molecule beginning in one secondary structure, $x$, will transition in the given time, $T$, to a target structure, $y$.  This is called the transition probability, and it is typically computed by either solving a system of linear differential equations or by computing a matrix exponential of a large matrix.  Here, we will use our method (denoted as TIPS) to approximate these transition probabilities.


{\bf Model.}  An RNA fold can be characterized by a set of base pairs, either C-G, A-U, or
G-U, each
of which specifies the sequence positions of the two bases involved in
the pairing.  We will default the discussion to RNA sequences where we
are interested in pseudo-knot-free RNA structures. These secondary
structures can be represented as a planar circle graph with the
sequence arrayed along a circle and non-crossing arcs between
positions of the sequence which are base paired. Here, we will use
structure to mean secondary structure.  The folding of a molecule into
secondary structures happens in a dynamic fashion.  


In the pathway model we consider, successive structures $X_i$ and $X_{i+1}$ must differ by exactly one base pair.  Let $X_1 = x$ and $X_N = y$ where $x$ is the given start structure and $y$ is the given final structure.  See for example Figure~5 of the Supplement, where a folding path is given for a short RNA (holding times not shown) with $x$ being the unfolded state and $y$ being the Minimum Free Energy~(MFE) structure.


To formalize the folding pathway, we need to introduce the generator matrix, $Q$.  This matrix contains an entry for every possible pair of secondary structures.  The Kawasaki rule gives the rate of the probabilistic process moving from structure $x$ to structure $x'$ as $\lambda(x) \nu(x,x') = \exp{(E(x) - E(x')) / (kT)}$ if $x' \in R(x)$, and zero otherwise 
where $E(x)$ is the energy of structure $x$, $R(x)$ is the set of secondary structures within one base pair of structure $x$ and $k$ is the Boltzmann constant.  When given a nucleic acid sequence of $m$ bases, there are at most $O(3^m)$ secondary structures that can be created from it,  making the size of the generator matrix exponential in the sequence length.  This model was described by Flamm et al.~\cite{Flamm2000}.





{\bf Results.} In this section, we compare the accuracy of the transition probability estimates given by our method (TIPS) to those obtained by forward sampling method (FS) which is still widely used in the field of RNA folding pathways~\cite{Flamm2000,Schaeffer2012Multistrand}. We used the RNA molecules shown in Supplement Table: Biological RNA Sequences.

For each method (TIPS and FS) and molecule, we first approximated the probability $\P_x(X_N = y)$ that beginning in its unfolded structure $x$, the molecule would end, after folding time $T$, in its MFE structure $y$.  We then computed, as a reference, the probability of this transition using an expensive matrix exponential. Computing the matrix exponential on the full state space was only possible for the RNAs of no more than 12 nucleotides. For the longer RNAs, we restricted the state space to  a connected subset $S$ of secondary structures~\cite{kirkpatrick_new_2013}.  While our method scales to longer RNAs, we wanted to be able to compare against forward sampling and to the true value obtained by matrix exponentiation.

We ran the experiments with a range of number of particles, $\{5^1, 5^2, \cdots, 5^6\}$, for 30 replicates on folding times from $\{0.125,0.25,\cdots,8\}$.
Here, similarly to the previous example, we compare the performance of the two methods by looking at the absolute log error of the estimate $\hat p$ (i.e., error($\hat p$)~$=|\log \hat p - \log \P_x(X_N = y)|$) over all replicates. 
The parameters used for the TIPS method are as follows: $\alpha = \frac{2}{3}$ and $\beta = \max(0.25, 1 -\frac{T}{16})$ where $T$ is the specified folding time interval. 

\begin{figure*}[t]
\begin{center}
	\subfloat[\tiny{RNA21- error}]{\label{fig: RNA21-a}\includegraphics[width=.45\textwidth,height=0.127\textheight]{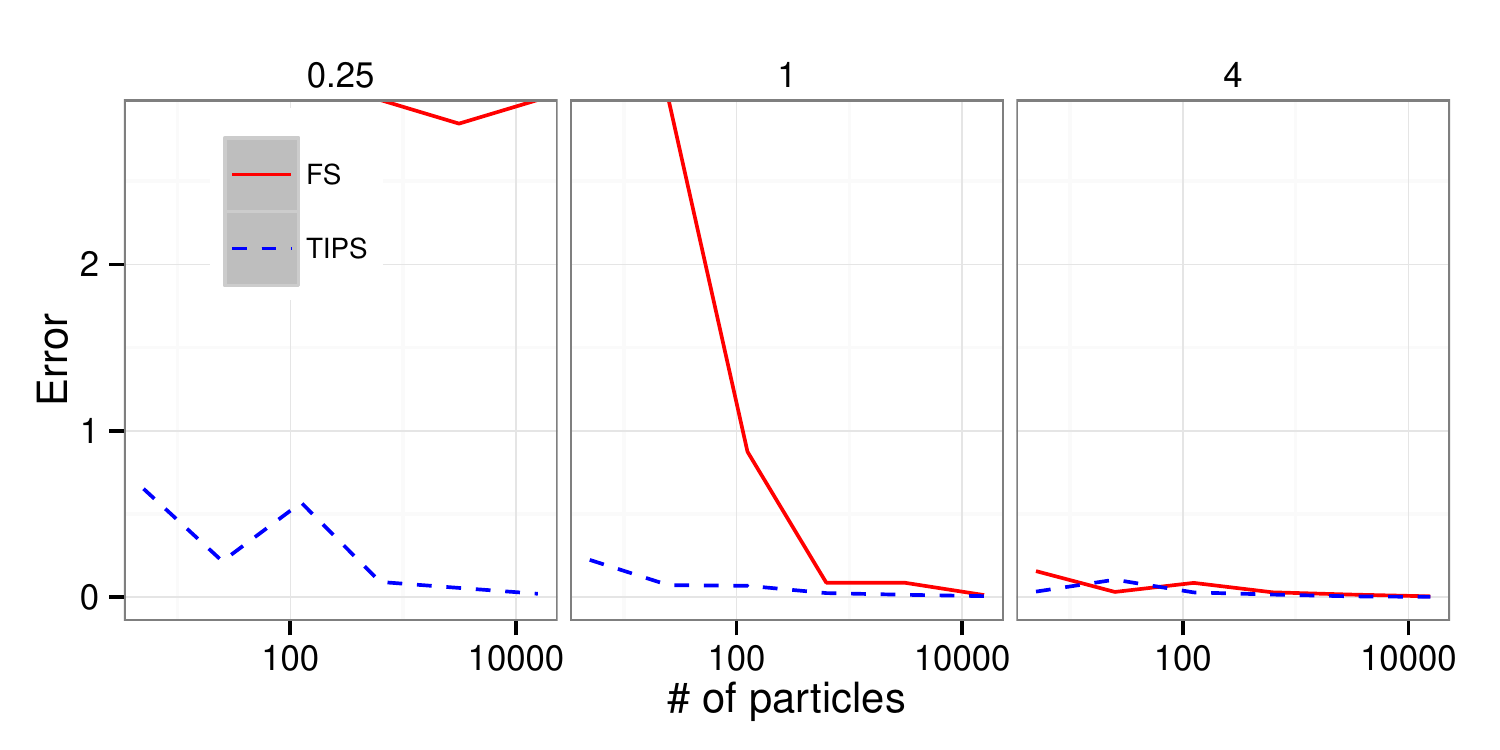} }
		\subfloat[\tiny{RNA21-~CPU time~(ms)}]{\label{fig: RNA21-b}\includegraphics[width=.18\textwidth,height=0.12\textheight]{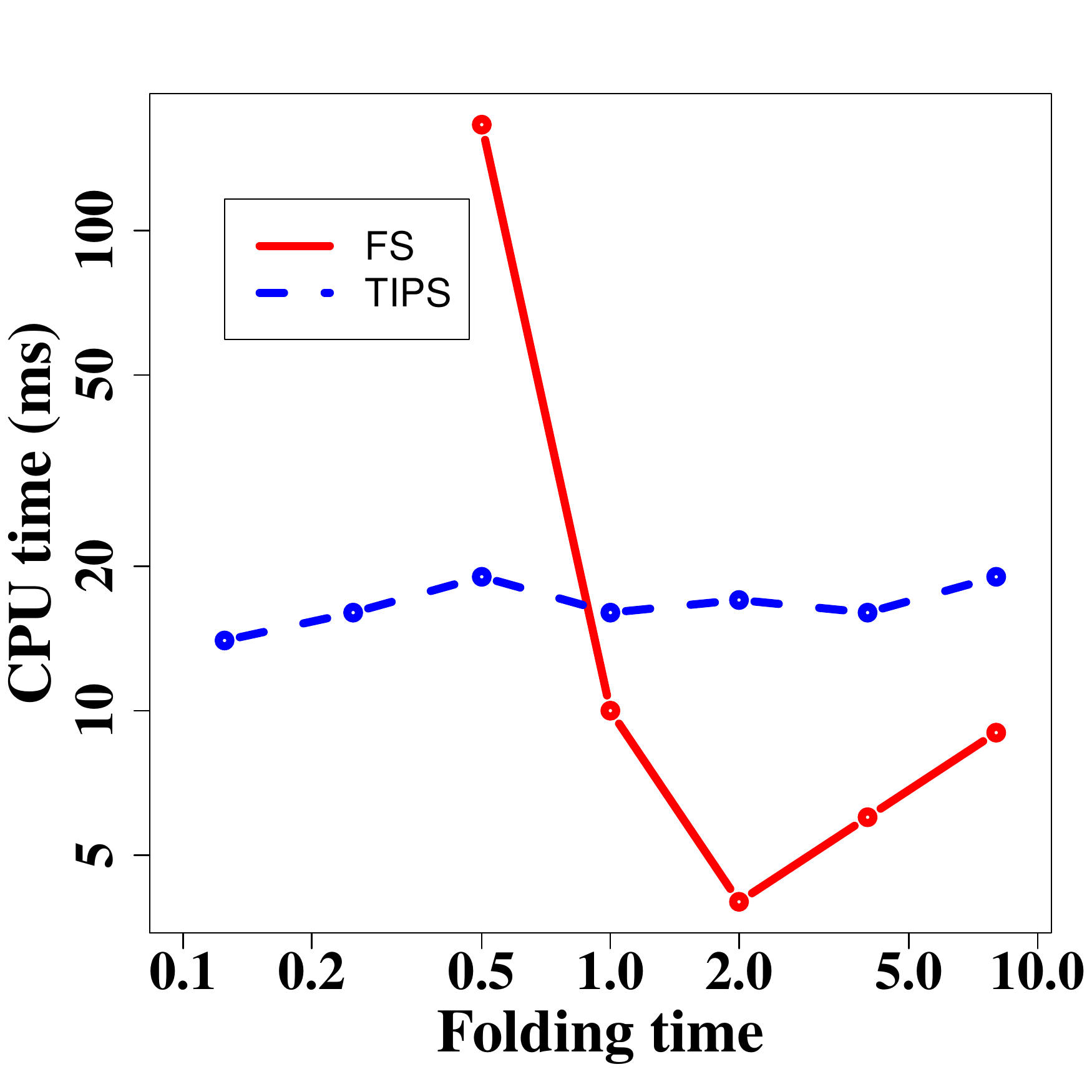} }
		\subfloat[\tiny{RNA21- variance}]{\label{fig: RNA21-c}\includegraphics[width=.18\textwidth,height=0.12\textheight]{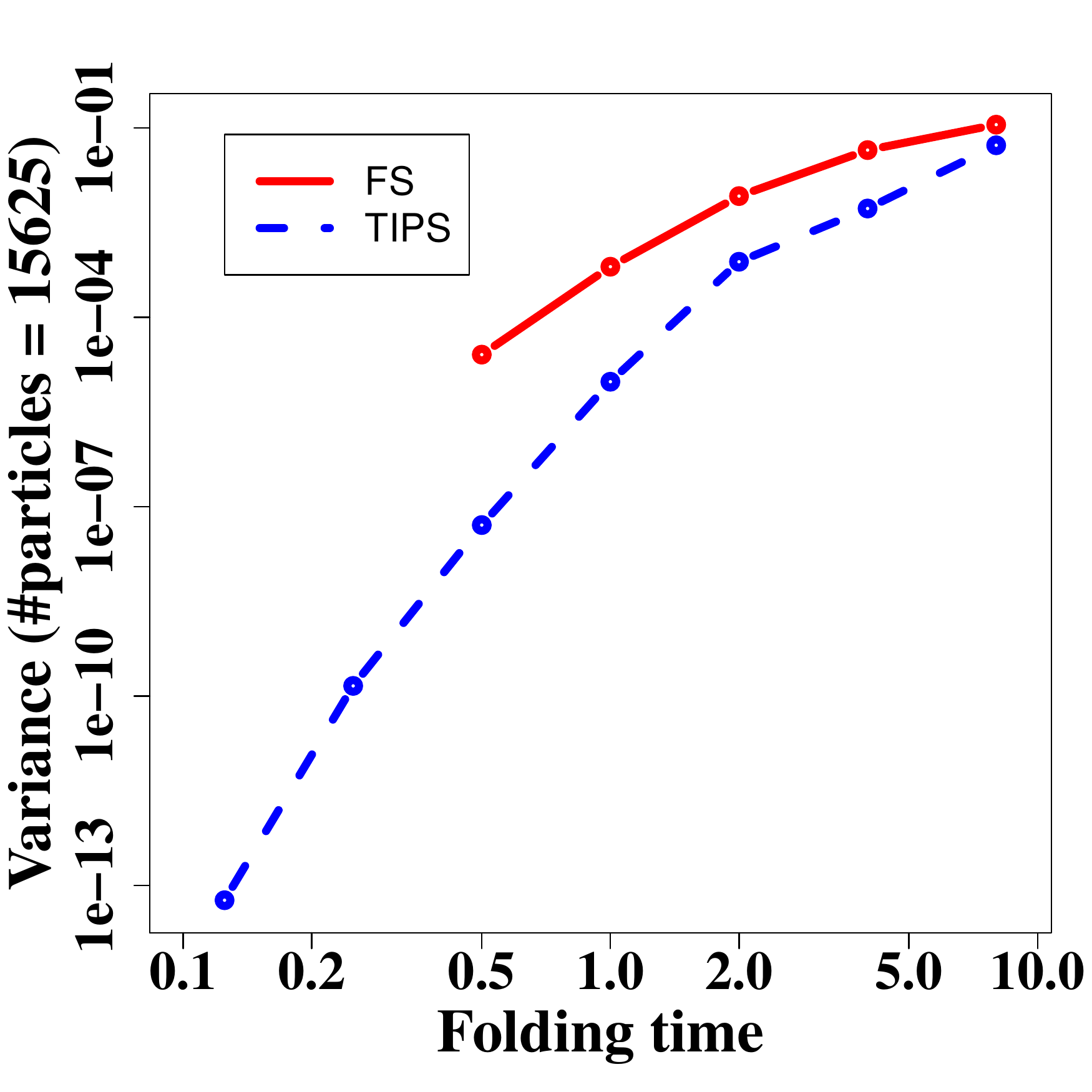}}
	\vspace{-0.4cm}	
\subfloat[\tiny{HIV23~- error}]{\label{fig: HIV-a}\includegraphics[width=.45\textwidth,height=0.127\textheight]{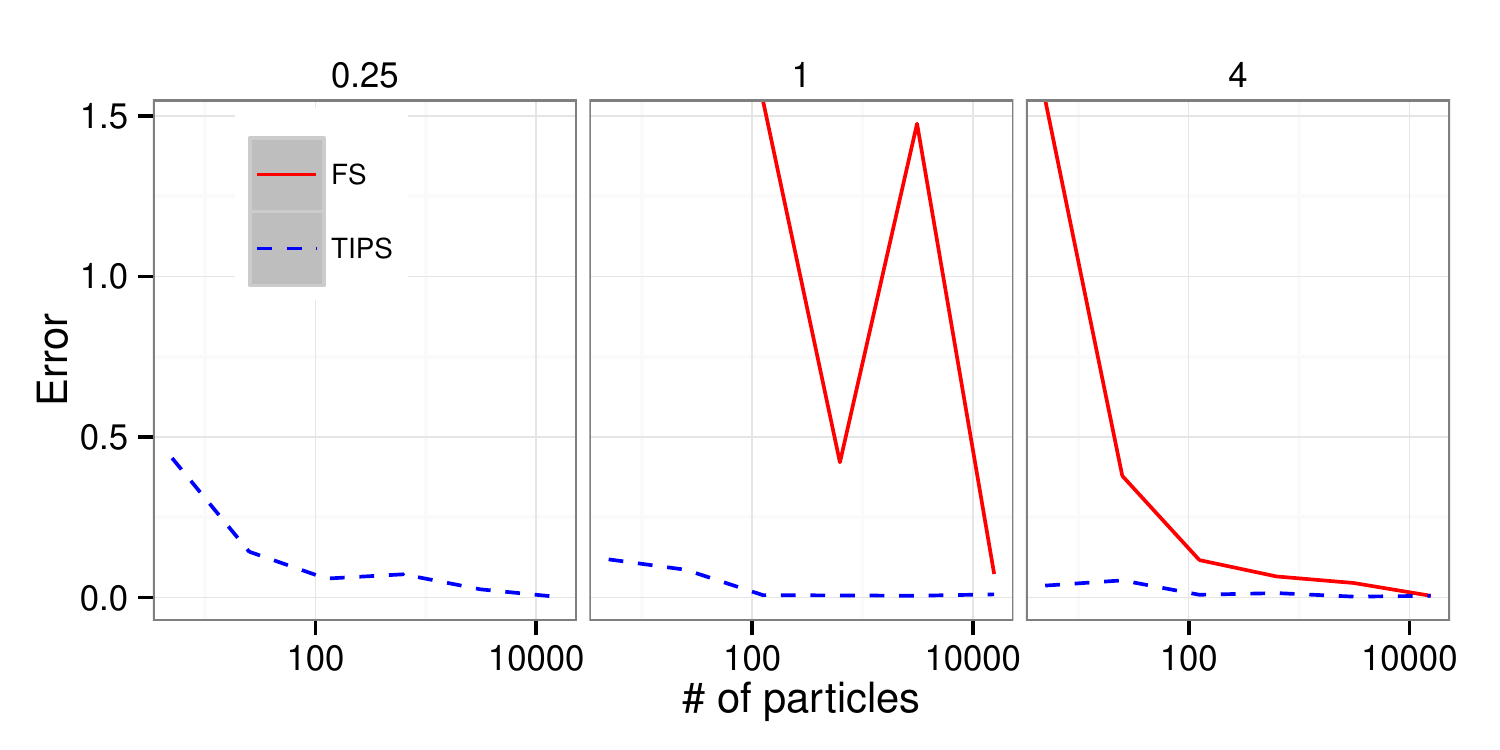} }
	\subfloat[\tiny{HIV23~- CPU time~(ms)}]{\label{fig: HIV-b}\includegraphics[width=.18\textwidth,height=0.12\textheight]{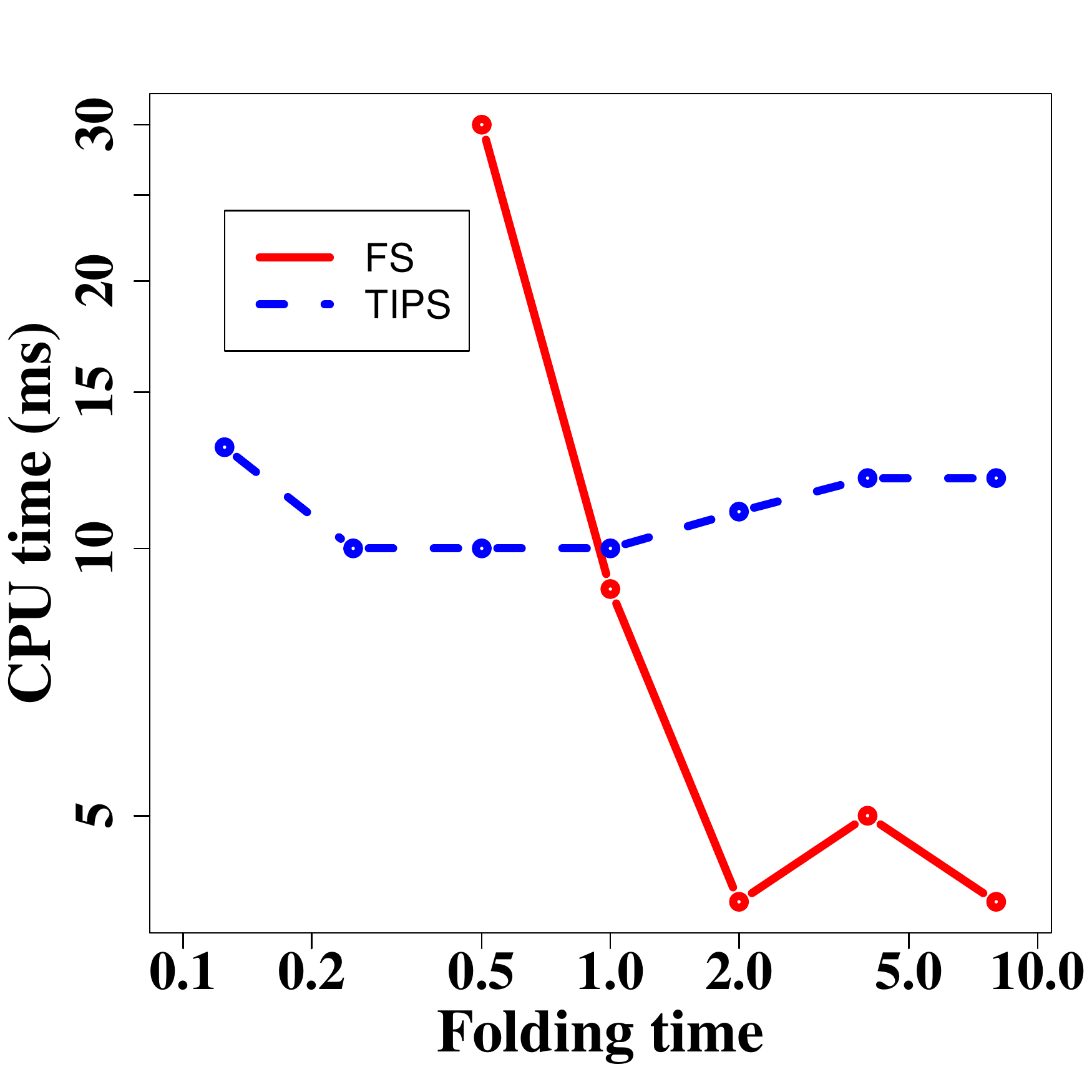} }
	\subfloat[\tiny{HIV23~- variance}]{\label{fig HIV-c}\includegraphics[width=.18\textwidth,height=0.12\textheight]{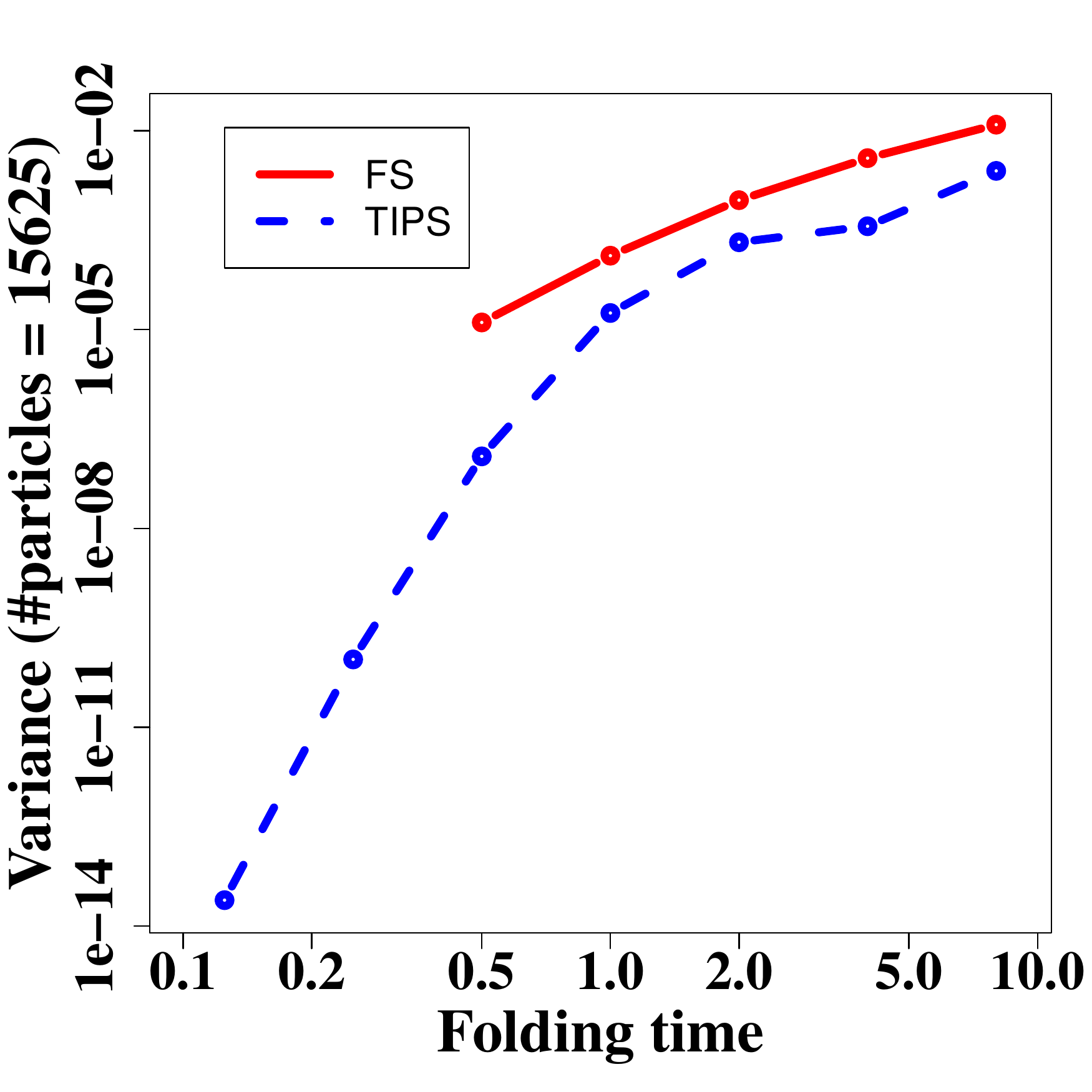}}

\caption[Performance vs. folding time]{\footnotesize{Performance of our method (TIPS) and forward sampling (FS) on RNA21 and HIV23 molecules with their \emph{subset} state space. The relative errors of the estimates vs. folding times, \{0.25,1,4\}, are shown (left) along with the CPU times corresponding to the minimum number of particles required to satisfy the accuracy level $I$ in milliseconds (middle) and the variance of TIPS and FS estimations (right) on folding times, $\{0.125, 0.25,\cdots, 8\}$.}}
\label{fig:RNAseq1}
\end{center}
\end{figure*}

Figures \ref{fig:RNAseq1}a, \ref{fig:RNAseq1}d show the performance of the FS and TIPS methods on selective folding times, $\{0.25,1,4\}$. Figures \ref{fig:RNAseq1}b, \ref{fig:RNAseq1}e show the CPU times (in milliseconds) corresponding to the minimum number of particles required to satisfy the certain accuracy level, $I=$~\{$\hat p:$ error($\hat p$)~$< 1.0$\} on all the folding times. Supplement Figure: Performance vs. folding time shows similar plots for two other RNA molecules.

The variances of FS and TIPS weights, for $5^6 = 15625$ particles, are also computed and compared on different folding times (see Figures \ref{fig:RNAseq1}c, \ref{fig:RNAseq1}f).

The graphs show that our novel method~(TIPS) outperforms FS in estimating the probability of transition from $x$ to $y$ in shorter folding times, since it needs many fewer particles (and correspondingly faster CPU times) than FS to be able to precisely estimate the probability. For instance, for the RNA21 molecule with folding time 0.25, FS cannot satisfy the accuracy level $I$, given above, even with 15625 particles, however TIPS only needs 5 particles with 16 ms of CPU time to satisfy the same accuracy level.
Similarly, the variance of our method is smaller by a larger margin (note that the variance is shown in log scale in Figures \ref{fig:RNAseq1}c, \ref{fig:RNAseq1}f).

For longer folding times in Figure \ref{fig:RNAseq1}, the performance of the TIPS and FS methods would be comparable (in terms of the obtained errors and CUP times) slightly in favour of forward sampling. For example, for the HIV23 molecule with folding time 4.0, TIPS and FS require 5 and 25 particles, and CPU times, 12 ms and 5 ms, respectively to satisfy $I$. 


One caveat of these results is that in contrast to the phylogenetic setup, where TIPS was not sensitive to a range of values of the tuning parameters $\alpha,\beta$, it was more sensitive to these tuning parameters in the RNA setup. See Supplement Figure: Tuning parameter $\alpha$. 
We believe that the behavior of our method is more sensitive to $\alpha,\beta$ in the RNA case because the sampled jump chains are typically longer.  
Intuitively, for longer folding times, the transition probabilities are more influenced by the low probability paths, as these low probability paths comprise a greater percent of all possible paths.  
This means that any setting of $\alpha$ that heavily biases the sampled paths to be from the region just around $x$ and $y$ will need to sample a large number of paths in order to approximate the contribution of paths with a low probability. 
This situation is analogous to the well-known problems in importance sampling of mismatches between the proposal and actual distributions.  Similar sampling considerations apply to parameter $\beta$ which controls the number of excursions from $y$.  If $\beta$ is too restrictive, again, paths will be sampled that do not well reflect the actual probability of excursions.
Parameter tuning is therefore an important area of future work.  It might be possible to use some automated tuners~\cite{Hutter2009,Wang2013AHMC} or to approach the problem by essentially creating mixtures of proposals each with its own tuning parameters.  

At the same time, note that the reason why FS can still perform reasonably well for longer folding times is that we picked the final end point to be the MFE, which has high probability under the stationary distribution. For low probability targets, FS will often fail to produce even a single hitting trajectory, whereas each trajectory sampled by our method will hit the target by construction.

\vspace{-0.1cm}
\section{Conclusion}
\label{sec:discussion}

We have presented an efficient method for approximating transition probabilities and posterior distributions over parameters in countably infinite CTMCs. We have demonstrated on real RNA molecules that our method is competitive with existing methods for estimating the transition probabilities which marginalize over folding pathways and provide a model for the kinetics of a single strand of RNA interacting chemically with itself. We have also shown, using a realistic, context-dependent indel evolutionary process, that the posterior distributions approximated by our method were accurate in this setting.

What makes our method particularly attractive in large or countably infinite state space CTMCs is that our method's running time per particle is independent of the size of the state space. The running time does depend cubically on the number of imputed jumps, so we expect that our method will be most effective when the typical number of transitions between two observations or imputed latent state is moderate (no more than approximately a thousand with current architectures). The distribution of the jump chain should also be reasonably concentrated to ensure that the sampler can proceed with a moderate number of particles. We have shown two realistic examples where these conditions are empirically met.

\subsection*{Acknowledgment}

This work was partially funded by an NSERC Discovery Grant and a Google Faculty Award. Computing was supported by WestGrid.

{\footnotesize
\bibliography{bib}
\bibliographystyle{icml2014}
}
\end{document}


\maketitle

This Supplement contains the proofs and pseudo-codes of the methods referenced in the {\bf Methodology} section of the paper, as well as supplemental figures and tables for the {\bf Numerical examples} section of the paper.  The proofs involve the basic properties of our CTMC approach, and provide rigorous justification for the algorithms presented in the paper. The pseudo-codes illustrate the overviews of the steps of different methods used in the manuscript. The supplemental figures and tables also provide more extensive justification for the practical applicability of our method to both the phylogenetic and RNA settings.

\section{Methodology}

We prove here the results stated in the methodology section of the main paper.

\subsection{Proposition 1}

\begin{proof} 
We have, for any $x^* = (x_1, x_2, \dots, x_n) \in{\mathcal X}^*$,
\begin{align*} 
\1(x_n = y)&\frac{ \left(\prod_{i=1}^{n-1} \nu(x_i, x_{i+1})\right) \P\left(\sum_{i=1}^{n-1} H_i \le T < \sum_{i=1}^n H_i\bigg|X^* = x^*\right)}{\P_x(X_N = y)}\\ &= \frac{\1(x_n = y) \P_x(X_1 = x_1, \dots, X_n = x_{n})}{\P_x(X_N = y)} \\ &  \ \ \ \  \ \  \times\P\left(\sum_{i=1}^{n-1} H_i \le T < \sum_{i=1}^n H_i\bigg|X_1 = x_1, \dots, X_n = x_n\right) \\
&= \1(x_n = y)\frac{1}{\P_x(X_N = y)} \\ &  \ \ \ \  \ \ \times \P_x\left(\sum_{i=1}^{n-1} H_i \le T < \sum_{i=1}^n H_i, X_1 = x_1, \dots, X_n = x_n \right) \\
&= \P_x(X^* = x^*| X_N = y).
\end{align*}Since the right hand side is a conditional distribution, 
\begin{align*} 
\pi(x^*) = \P_x(X^* = x^*| X_N = y),
\end{align*}is indeed a normalized probability mass function.
\end{proof}

\subsection{Proposal distributions}

We show that the proposal defined in Equation~(2) of the main paper hits the target end point $y$ with probability one under the following assumptions:
\begin{enumerate} 
  \item The potential $\rho^y(x)$ takes the value zero if and only if $x = y$.
  \item The potential always changes by one in absolute value for all proposed states:
  \begin{align*} 
  \tilde \P_x\left(|\rho^y(X_2) - \rho^y(x) | = 1 \right) = 1.
\end{align*}
   \item For all states $x\neq y$, there is always a way to propose a state that results in a decrease in potential:
   \begin{align*} 
   \tilde \P_x(\rho^y(X_2) < \rho^y(x) ) > 0\ \ \ \ \textrm{for all }x\in\states, x\neq y.
\end{align*} 
\end{enumerate}
To simplify the notation, we will drop the $y$ superscript for the remaining of this section.

To prove that the process always hits $y$, it is enough to show that the sequence $\rho(X_n)$ is a supermartingale, which in our case reduces to showing that $\E[\rho(X_2) | X_1] \le \rho(X_1).$

Note that the last condition ensures that the normalizer $\sum_{x'_{2} \in D(X_1)}  \nu(X_1,x'_{2})$ is always positive, hence our expression of the proposal is always well defined. Note that technically, we should also require $\P_x(\rho^y(X_2) < \rho^y(x) ) > 0$ to ensure that the second normalizer, $\sum_{x'_{2} \notin D(X_1)} \nu(X_1,x'_{2})$, is also positive, but if this is not the case, the proposal can always be replaced by $\nu$ in these cases without changing the conclusion of the result proven here.

Using the second condition, we have:
\begin{align*} 
\E[\rho(X_2) | X_1] &= \alpha_{X_1}(\rho(X_1) - 1) + (1- \alpha_{X_1})(\rho(X_1) + 1) \\
&= 1 - 2 \alpha_{X_1} + \rho(X_1) \\
&\le \rho(X_1).
\end{align*}

Finally, since the supermartingale $\rho(X_n)$ is non-negative, $\tilde \P(N < \infty) = 1$, we conclude that the process always hits $y$.


\subsection{Proposition 2}

Let $\check X_1, \check X_2, \dots$ and $\check H_1, \check H_2, \dots$ denote the states and holding times respectively of a CTMC with rate matrix $\check Q$. The states take values in $\{1, 2, \dots, n+1\}$, and we let $\check \P_1$ denote the path probabilities under this process conditioned on starting at $X_1 = 1$. Let $\check N$ be defined similarly to $N$ (the random number of states visited): 
\begin{align*}
(\check N = n) &= \left(\sum_{i=1}^{n-1} \check H_i \le T < \sum_{i=1}^n \check H_i\right) \\
&= \left\{\omega \in \check \Omega : \sum_{i=1}^{n-1} \check H_i(\omega) \le T < \sum_{i=1}^n \check H_i(\omega) \right\}.
\end{align*}Here, $\check \Omega$ is an auxiliary probability space used to define the above random variables:
\begin{align*} 
\check X_i &: \check \Omega \to \states \\
\check H_i &: \check \Omega \to [0, \infty).
\end{align*}

\begin{proof} 
For all $i\in\{2, \dots, n+1\}$, only state $i-1$ has a positive rate of transitioning to state $i$, therefore $(\check{X}_{i} = j) \subset (\check{X}_{i-1} = j-1)$ for all $j$. Applying this inductively yields:
\begin{align*} 
\left(\exp(T \check Q)\right)_{1,n} &= \check \P_1\left(\check X_{\check N} = n\right) \\
&= \check \P_1\left(\check X_{\check N} = n, \check X_{\check N-1} = n-1\right) \\
&\ \ \vdots \\
&= \check \P_1\left(\check X_{\check N} = n, \check X_{\check N-1} = n-1, \dots, \check X_{1} = n-\check N+1\right) \\
&= \check \P_1\left(\check N = n, \check X_1 = 1, \check X_2 = 2, \dots, \check X_n = n\right) \\
&= \check \P_1\left(\check N = n\right) \check \P_1\left(\check X_1 = 1, \check X_2 = 2, \dots, \check X_n = n | \check N = n\right) \\
&= \check \P_1\left(\check N = n\right) \prod_{i=2}^n \check \P(\check X_i = i | \check X_{i-1} = i-1) \\
&= \check \P_1\left(\check N = n\right)  \\
&= \int \int \cdots \int_{h_i > 0 : h_1+h_2+\cdots+h_{n} = T} g(h_1, h_2, \dots, h_{n}) \ud h_1 \ud h_2 \dots \ud h_{n}.
\end{align*}
\end{proof}

In this part, for further clarity, we give the pseudo-codes of the proposed algorithms.\\

Our novel method (denoted as Time Integrated Path Sampling, TIPS) is demonstrated in Algorithm \ref{alg:TIPS}. This method uses the propose method introduced in Algorithm \ref{alg:prop} in order to sample each particle, consisting of a sequence of states starting at $x$ and ending at the target, $y$. The propose method also employs Algorithm \ref{alg:prop-hit} as a part of its structure to sample the particles hitting the target.  


\begin{algorithm}[!h]
\caption{: {\bf TIPS}($x, y, T$)}
\begin{algorithmic}
\STATE $s \gets 0$
\FOR{$k = 1, 2, \dots, K$}
\STATE $(L, \tilde p, p) \gets $~{\bf propose}$(x, \{y\})$
\STATE $\check{Q} \gets \check{Q}(L)$ \COMMENT{See Section `Analytic jump integration'}
\STATE $n = |L|$ \COMMENT{The length of the list of states $L$}
\STATE $s \gets s + p\times(\exp(T \check{Q}))_{1,n} / \tilde p$
\ENDFOR
\RETURN $s/K$
\end{algorithmic}
\label{alg:TIPS}
\end{algorithm}

\begin{algorithm}
\caption{: {\bf propose}($x, A$)}
\begin{algorithmic}
\STATE $(L, \tilde p, p) \gets ${\bf proposeHittingPath}($x, A, $ false)
\STATE $n \sim $~Geo$(\cdot, \beta)$ \COMMENT{Geometric with support 1, 2, \dots}
\STATE $\tilde p \gets p\ \times$~Geo$(n; \beta)$ \COMMENT{Multiply by geometric probability mass function}
\FOR{$i = 2, 3, \dots, n$}
\STATE $x' \gets$ {\bf last}$(L)$ \COMMENT{Last state visited in the list $L$}
\STATE $(L', \tilde p', p') \gets${\bf proposeHittingPath}($x', A, $ true)
\STATE $\tilde p \gets \tilde p \times \tilde p'$
\STATE $p \gets p \times p'$
\STATE $L \gets L \circ L'$ \COMMENT{Concatenation of the two lists}
\ENDFOR
\RETURN $(L, \tilde p, p)$
\end{algorithmic}
\label{alg:prop}
\end{algorithm}

\begin{algorithm}
\caption{: {\bf proposeHittingPath}($x, A, b$)}
\begin{algorithmic}
\STATE $p \gets 1$
\STATE $\tilde p \gets 1$
\STATE $L \gets$~{\bf list}($x$)  \COMMENT{Creates a new list containing the point $x$}
\FOR{$i = 1, 2, \dots$}
\IF{$x \in A$~and~(not($b$)~or~$i>1$)}
\RETURN $(L, \tilde p, p)$
\ENDIF
\STATE $x' | x \sim \tilde\P(\cdot|X_{i-1} = x)$
\STATE $\tilde p \gets \tilde p \times \tilde\P(X_i = x' | X_{i-1} = x)$
\STATE $p \gets p \times \nu(x, x')$
\STATE $L \gets L \circ x'$
\STATE $x \gets x'$
\ENDFOR
\end{algorithmic}
\label{alg:prop-hit}
\end{algorithm}

An overview of the parameter estimation method explained in the manuscript is also shown in Algorithm \ref{alg:PS}. In this algorithm, statio() computes the stationary probability mass function, for example Poisson in the Poisson Indel Process example \cite{Bouchard2012Evolutionary}.

\begin{algorithm}[!htbp]
\caption{: {\bf TIPS-parameters}($\{x_i, y_i, T_i\}$)}
\begin{algorithmic}
\STATE Initialization:
\bindent
\STATE Choose initial parameter $\theta^{(1)}$
\STATE  $z \gets 0$
\eindent
\FOR{$(t=2, \cdots,  T)$}
\STATE $\theta^* \sim q(\cdot|\theta^{(t-1)})$
\STATE $z^* \gets 1$
\FORALL{data index $i$}
\STATE $z^* \gets z^* \times$~{\bf TIPS}($x_i, y_i, T_i$)~$\times$~{\bf statio}($x_i$) 
\ENDFOR
\STATE $r \gets \frac{p(\theta^*)}{p(\theta^{(t-1)})} \ \frac{z^*}{z} \ \frac{q(\theta^{(t-1)} | \theta^*)}{q( \theta^* | \theta^{(t-1)})}$
\STATE Sample $u \sim \textrm{Uniform}(0,1)$
\IF {$u < \min\{1, r\}$} 
\STATE $z \gets z^*$
\STATE $\theta^{(t)} \gets \theta^{*}$
\ELSE
\STATE $\theta^{(t)} \gets \theta^{(t-1)}$
\ENDIF
\ENDFOR
\end{algorithmic}
\label{alg:PS}
\end{algorithm}

Moreover, Algorithm \ref{alg:SMCseq}, extends \cite{Saeedi2011Priors} and demonstrate the revised sequential Monte Carlo (SMC) method for approaching more general types of observations, for example a series of partially observed states, or a phylogenetic  tree with observed  leaves. Note that this algorithm is amenable to parallelization \cite{Lee2010,Jun2012Entangled}.

\begin{algorithm}[!htbp]
\caption{: {\bf TIPS-SMC}($A_i, T_i)$}
\begin{algorithmic}
\STATE Initialization:
\bindent
\STATE $w_{0,k} \gets 1/K$
\STATE $x_{0,k} \gets$~nil
\eindent
\FOR{$g = 1, 2, \dots, G$}
\FOR{$k = 1, 2, \dots, K$}
\STATE $(L, \tilde p, p) \gets$~{\bf propose}$(x_{g-1,k}, A_g)$
\STATE $\check{Q} \gets \check{Q}(L)$ \COMMENT{See Section `Analytic jump integration'}
\STATE $n = |L|$ \COMMENT{The length of the list of states $L$}
\STATE $w_{g,k} \gets w_{g-1,k} \times p \times (\exp(T_g \check{Q}))_{1,n} / \tilde p$
\STATE $x_{g,k} \gets$~{\bf last}$(L)$
\ENDFOR
\IF{ESS$(w_{g,\cdot}) <$~threshold}
\STATE $(w_{g,\cdot}, x_{g,\cdot}) \gets$~{\bf resample}$(w_{g,\cdot}, x_{g,\cdot})$ \COMMENT{Perform SMC resampling step}
\ENDIF
\ENDFOR
\end{algorithmic}
\label{alg:SMCseq}
\end{algorithm}

\section{Numerical examples}

In this section we have figures and tables for both the phylogenetic and RNA settings.  These give more detail on the results that we have obtained from applying our method to these CTMCs.

\subsection{Phylogenetics}

\subsubsection{Validation}

\begin{figure}[!htbp]
\begin{center}
\begin{tabular}{ccccc}
{\footnotesize Exact} & {\footnotesize TIPS-parameters} & {\footnotesize Samples} & {\footnotesize Monte Carlo Quartiles} &\\
\includegraphics[width=1.2 in]{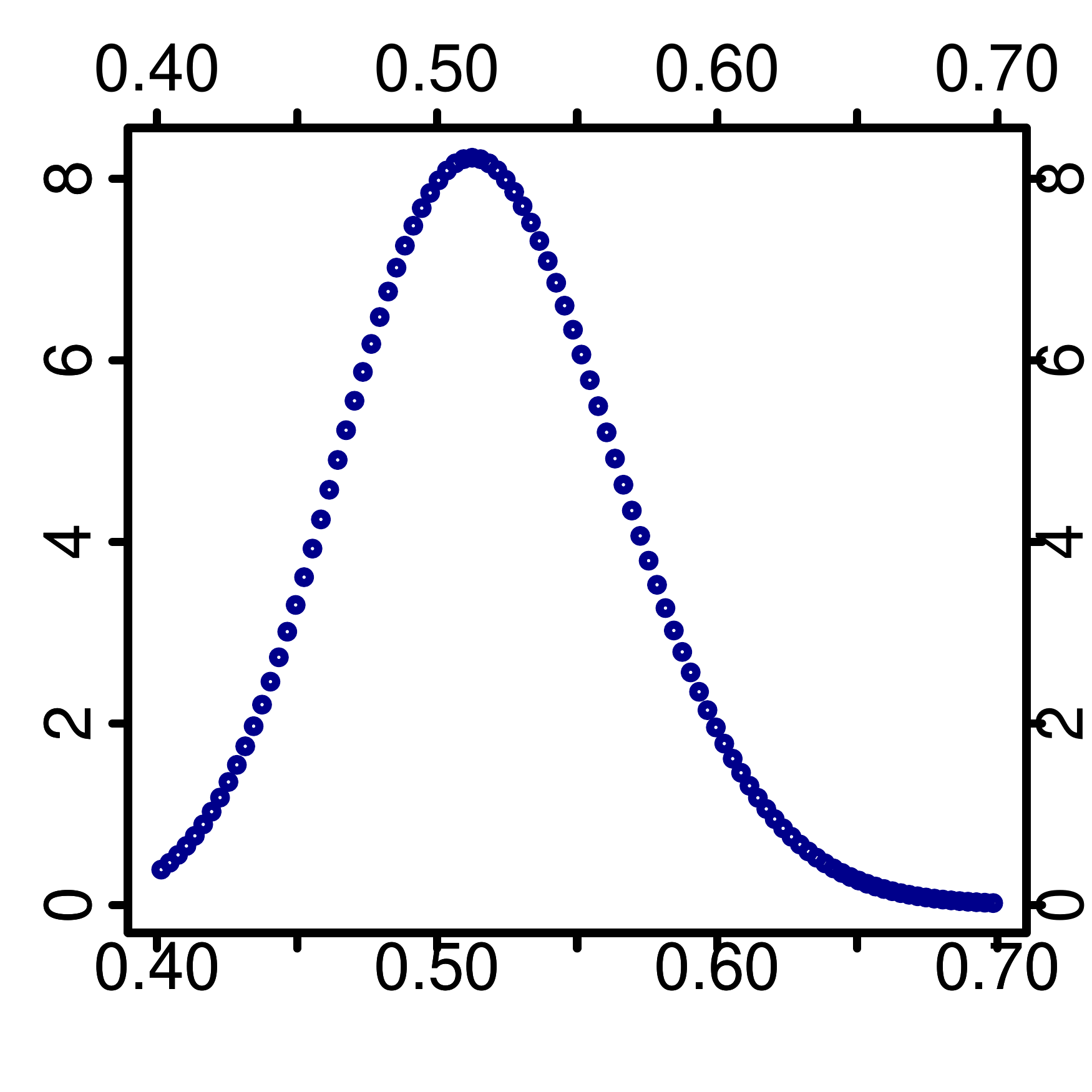}     & \includegraphics[width=1.2 in]{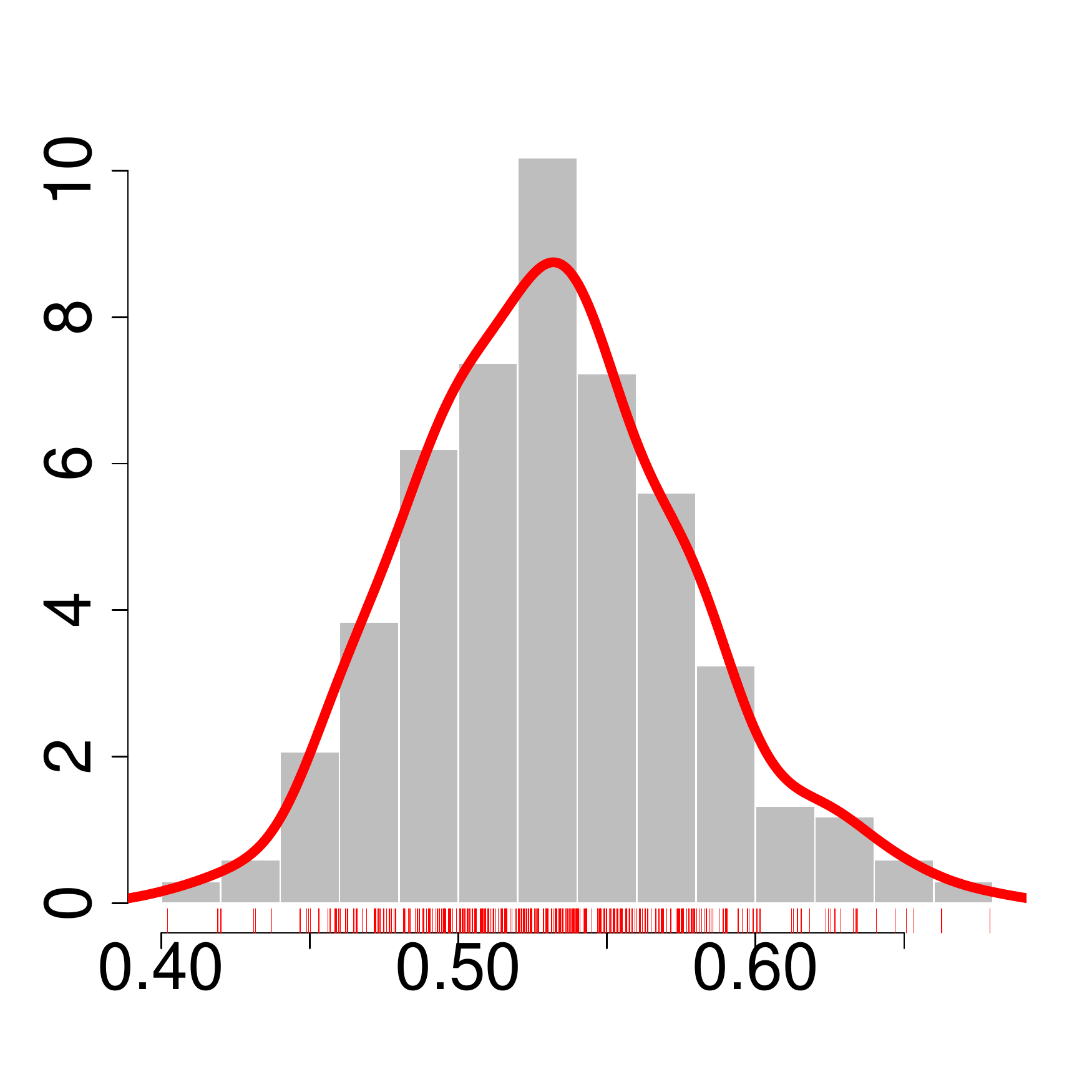} & \includegraphics[width=1.2in]{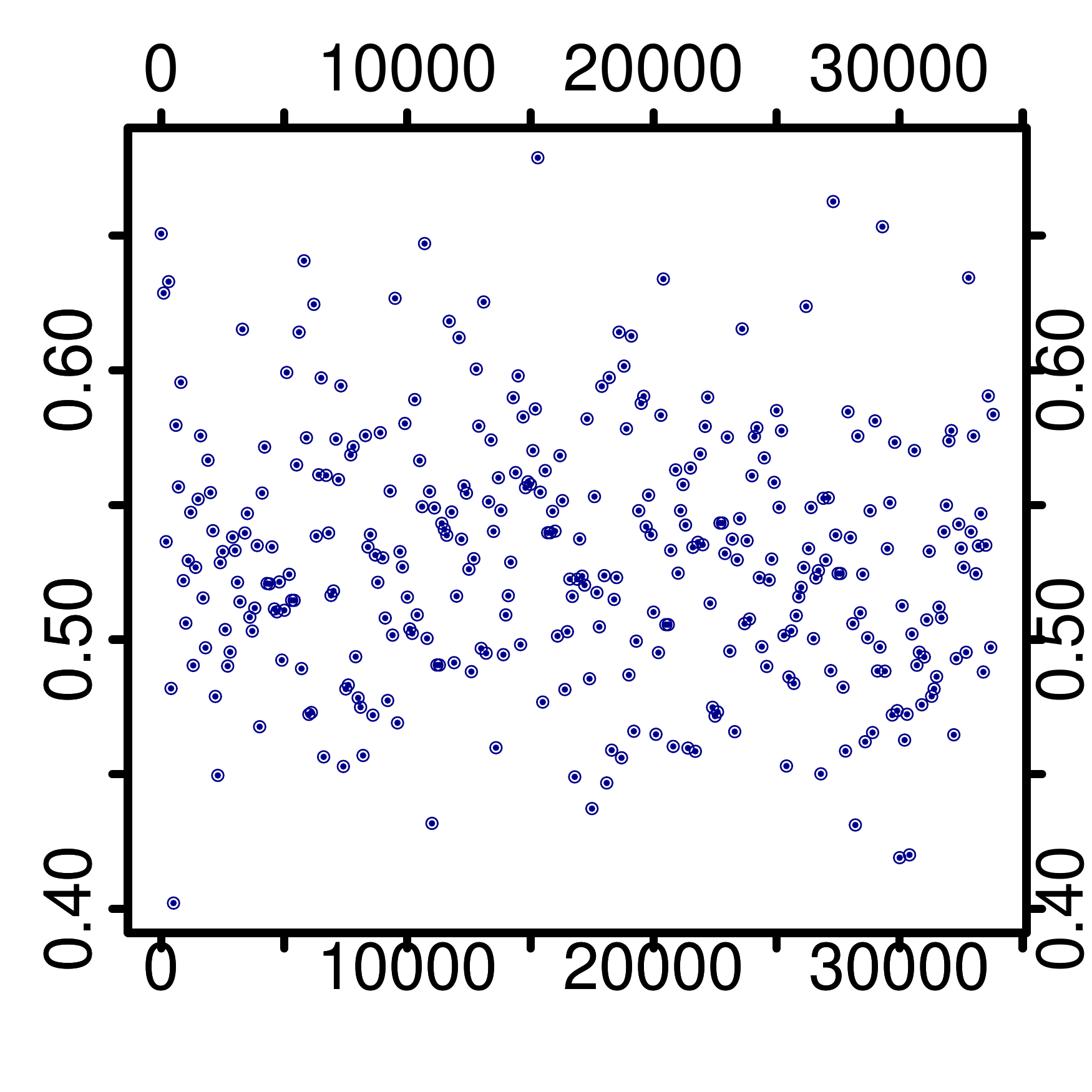} & \includegraphics[width=1.2in]{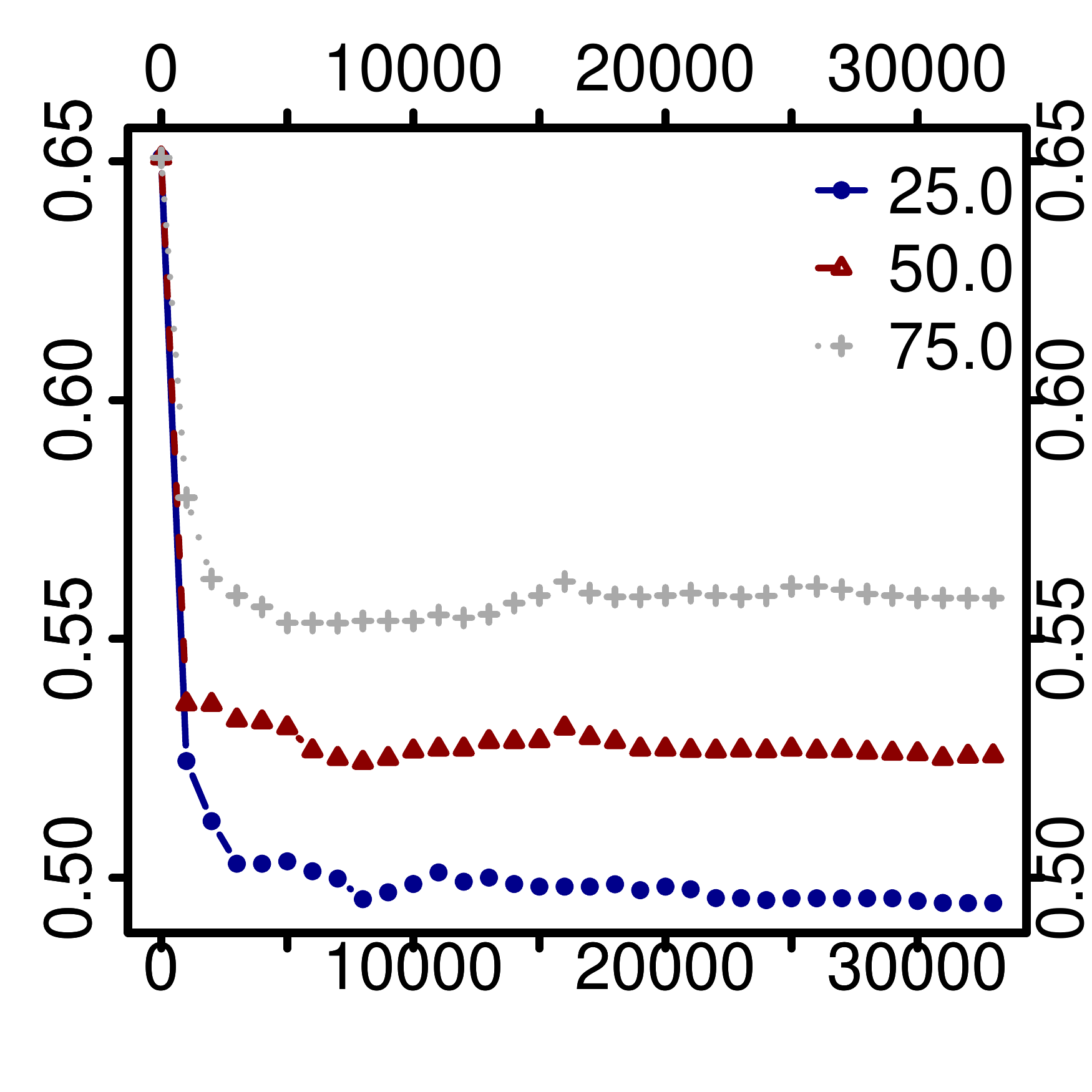} \\
\includegraphics[width=1.2 in]{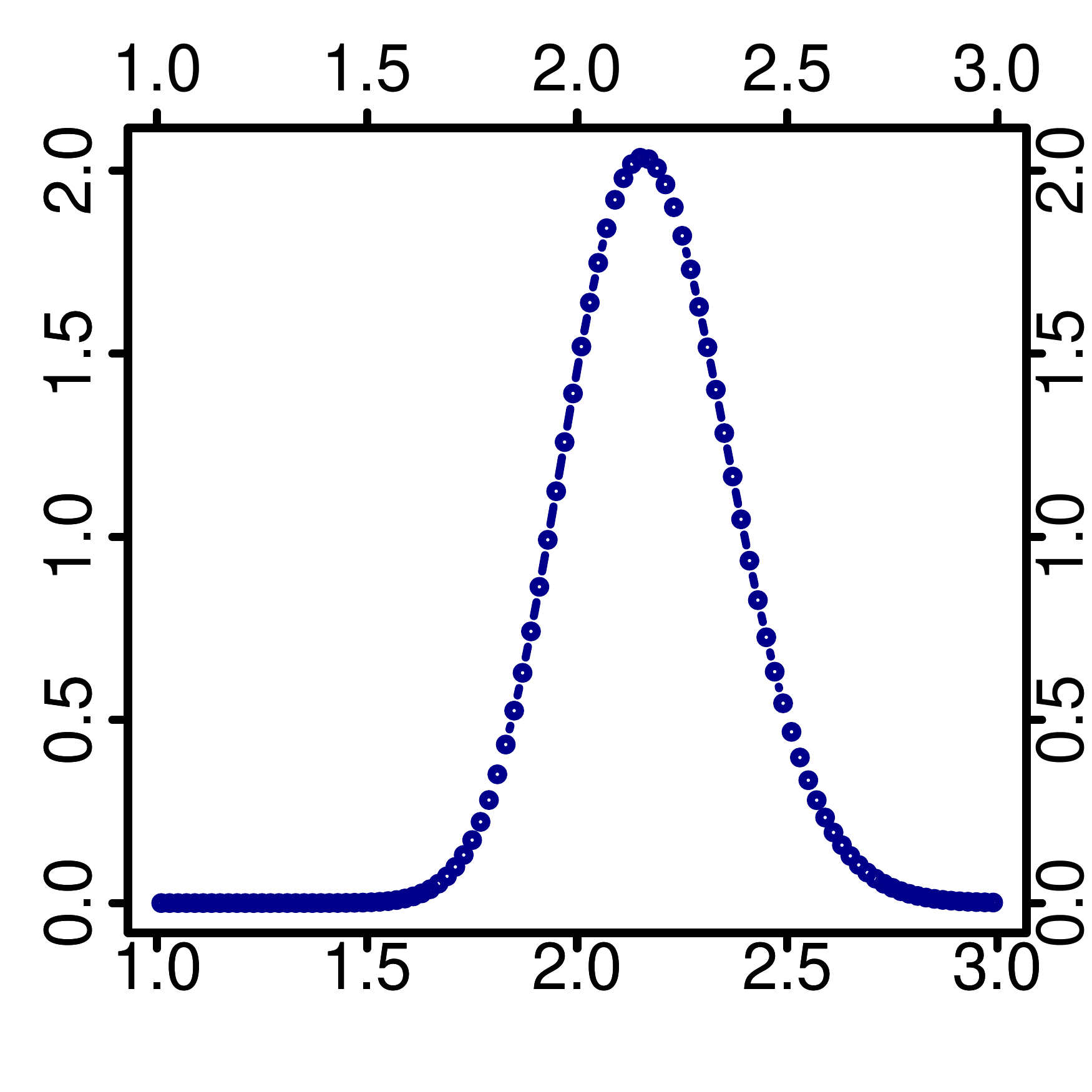}     & \includegraphics[width=1.2 in]{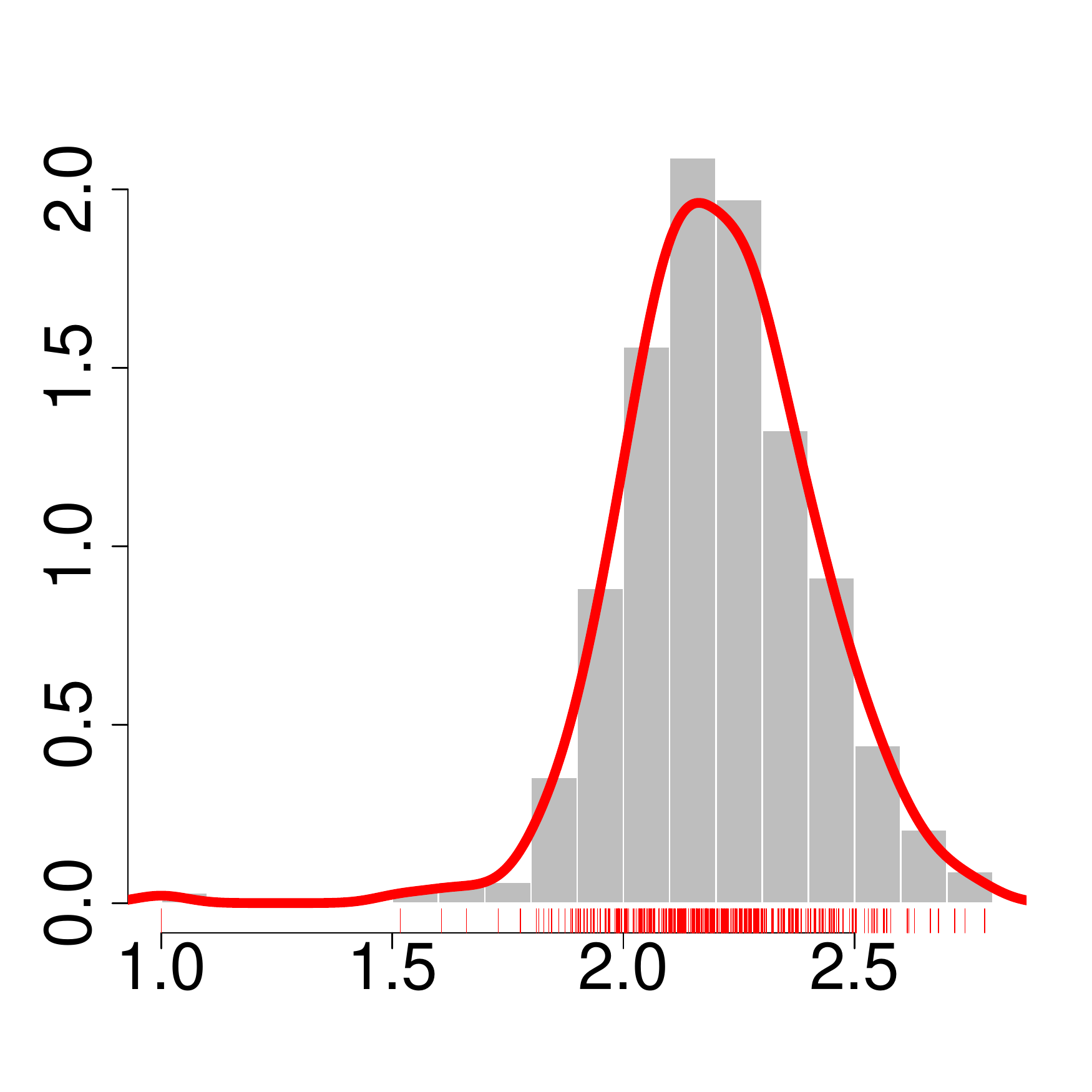} & \includegraphics[width=1.2in]{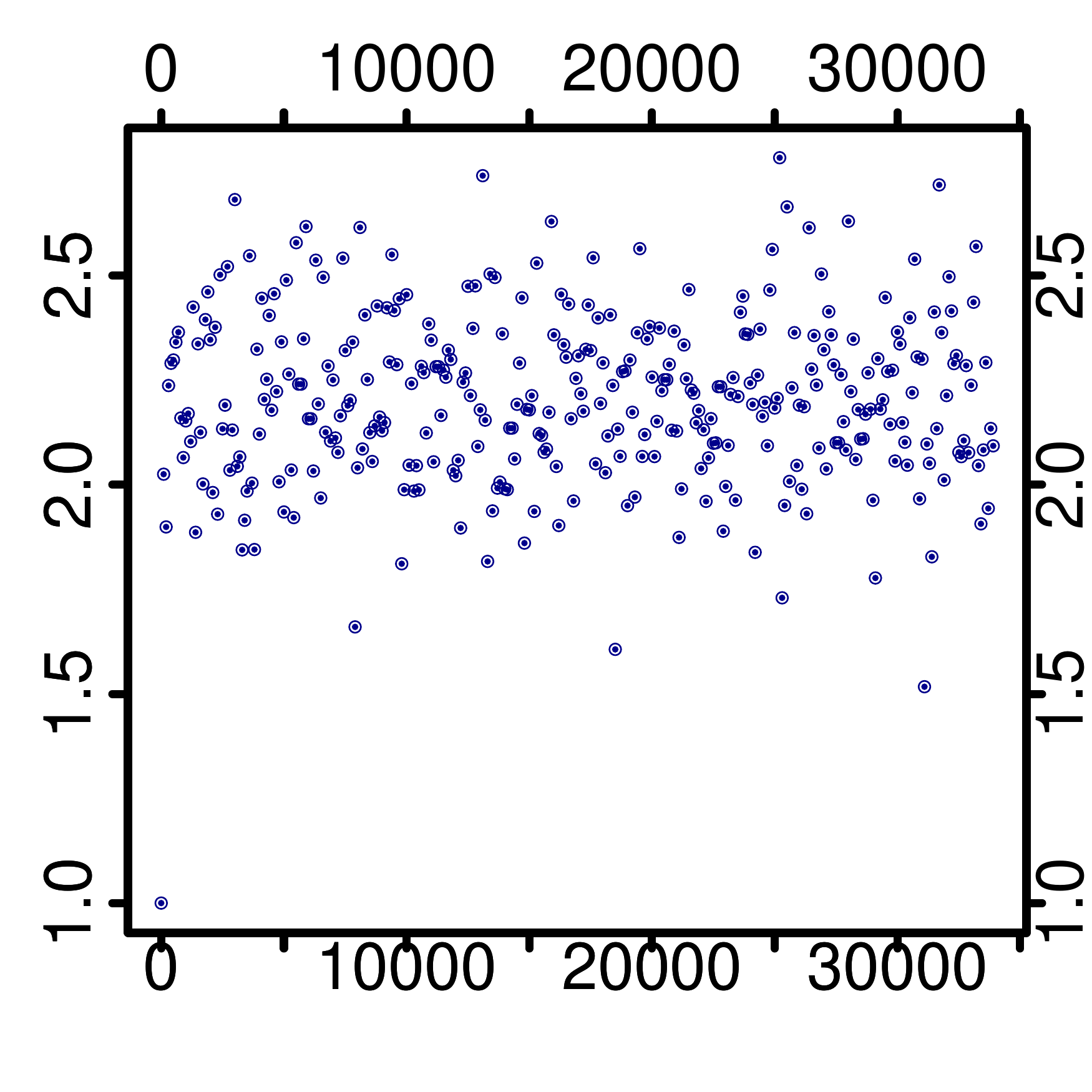} & \includegraphics[width=1.2in]{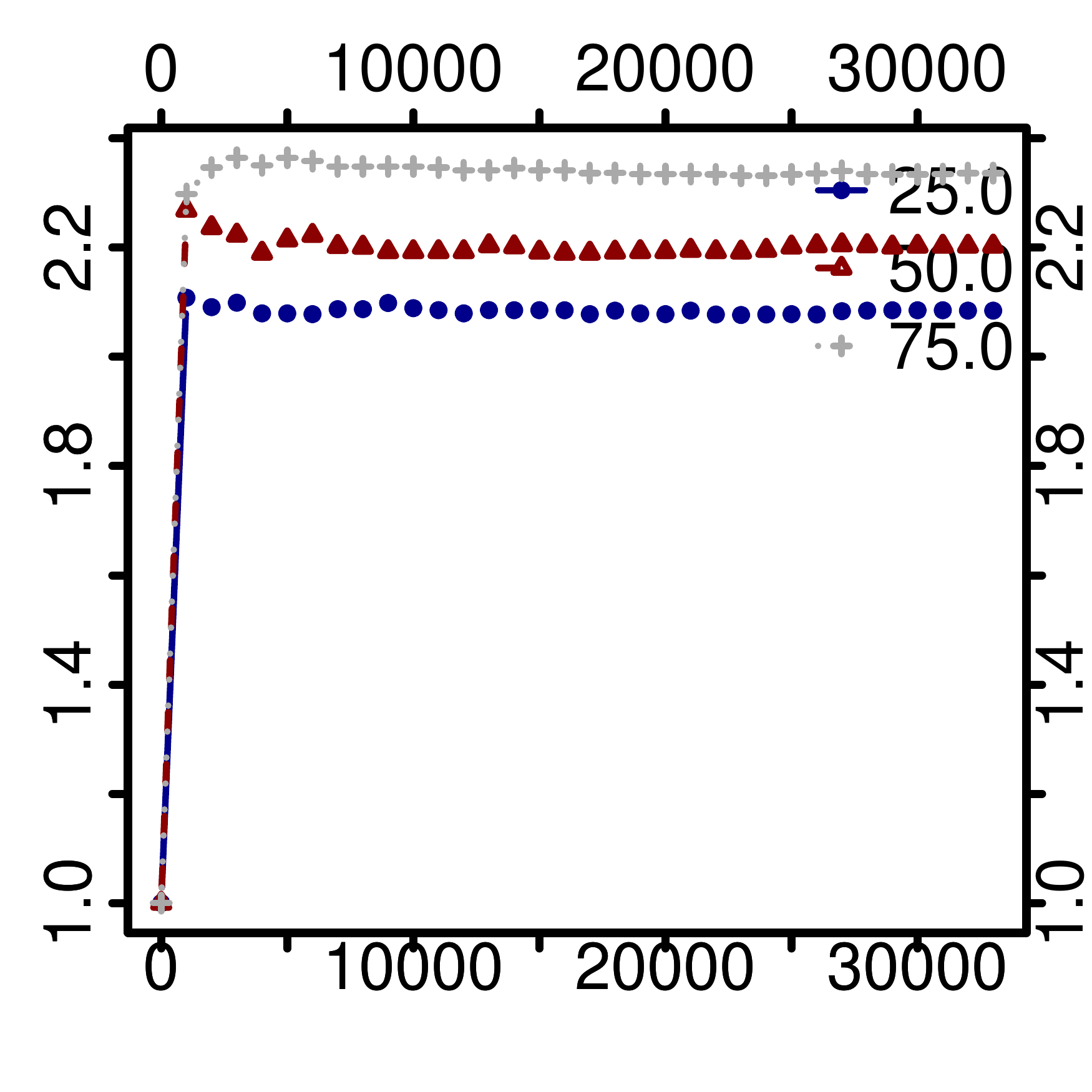} \\
\end{tabular}
\end{center}
  \caption{{\bf Results of parameter $\mu$} ($1^{st}$ row) in the validation of the implementation using the Poisson Indel Process
(PIP). From left to right:
posterior obtained via numerical methods; approximate posterior using our method 
with 64 particles and 35,000 GIMH iterations; sampled parameter values; convergence
of the quantile computed from the GIMH output. For further clarity, the results of parameter $\lambda$ ($2^{nd}$ row) are also shown.
}
  \label{fig:pipMuRestuls}
\end{figure}

We generated 200 pair-wise alignments with $T = 3/10, \lambda = \lambda_\textrm{pt} = 2, \mu = \mu_\textrm{pt} = 1/2$ and held out the mutations and the true value of parameters $\lambda$ and $\mu$. We approximate the posterior using our method.  We show the results of $\lambda$ in the paper and $\mu$ in Figure~\ref{fig:pipMuRestuls}.  In both cases the posterior approximation is shown to closely mirror the numerical approximation. The evolution of the Monte Carlo quartiles computed on the prefixes of Monte Carlo samples also show that the convergence is rapid in this case.

\begin{table}
\begin{center}
\begin{tabular}{crrr}
\toprule
& & \multicolumn{2}{c}{ESS} \\
\cmidrule(l){3-4}
Parameter & N. Particles & FS & TIPS \\
\midrule
$\lambda$&10&175.9&1,556.2 \\
&100&44.6&{\bf 7,082.7}\\
&1000&25.6&927.8\\
&10000&35.2&147.5\\
&100000&12.2&12.4\\
\midrule
$\mu$&10&11.3&774.1\\
&100&90.2&{\bf 6,761.2}\\
&1000&31.0&903.5\\
&10000&48.3&128.9\\
&100000&16.7&12.0\\
\bottomrule
\end{tabular}
\end{center}
\caption{{\bf Varying the number of particles per MCMC step.} We show the effective sample size obtained in a fixed computational budget (wall clock time of 3 days), comparing a GIMH based on our method (TIPS), compared to the forward sampling method (FS). Refer to Section~3 of the main paper for details.}
\label{table:results}
\end{table}

\subsubsection{Tree inference} 

We start by introducing some notation for data on a phylogenetic tree $\tau$. Let $v_0$ denote the root, and  $V(\tau)$  the other nodes. Let $\varrho(v)$ denote the parent of  the node $v\in V(\tau)$. 
Let  $X^{*(v)}=(X^{(v)}_1, \cdots, X^{(v)}_{N_v})$ denote  the sequence of  molecular strings that evolve from  $\varrho(v)$ to $v$.  The  
string $X^{(v)}_{N_v}$ at Node $v$ is also denoted by $X^{(v)}$ for simplicity. Note that only the strings at the leaves are observed, denoted  $\observations$. 
 Denote all unobserved  strings by $X^{*}(\tau)=\{X^{*(v)}: v\in V(\tau)\}\backslash \observations$, where $\backslash$ is the  set difference  symbol. The probability of  $\observations$  and $X^{*}(\tau) = x^{*}(\tau)$  given $\tau, \theta$ is {\footnotesize 
\begin{flalign*}
&\P_\theta\bigg(\observations, X^{*}(\tau) = x^{*}(\tau)\bigg|\tau\bigg) =\\
&  \P\left(X^{*(v_0)} = \str^{*(v_0)}\right) \prod_{v \in
V(\tau)} \P_{x^{(\varrho(v))},\theta}\left(X^{*(v)} = x^{*(v)}\right).
\end{flalign*}
}
We use an improper uniform distribution over the strings  as the 
distribution for the root sequence.

In the  Bayesian framework, we aim at  the posterior on
$\tau$, $X^{*}(\tau)$, $\theta$ given $\observations$, which has a density  
proportional to
$\gamma(\tau, \str^{*}(\tau), \theta)  = \P_\theta(\observations, X^{*}(\tau) = x^{*}(\tau)|\tau)p(\tau)p(\theta)$, 
where  $p(\tau)$ is a prior for $\tau$.  

\newcommand\leftsub{\textrm{left}}
\newcommand\rightsub{\textrm{right}}
\newcommand\treesub{\textrm{tree}}

For fixed evolutionary parameters, we use the framework of \cite{Wang2012PhDThesis} 
to estimate the posterior of $\tau$. In this framework, we let 
the $r$-th partial state $\pstate_r$ be a forest that includes the forest topology and the
associated branch lengths, denoted  $\tau_r$,  as well as the unobserved strings at the root of each tree in that forest, 
 $\str^{*}(\tau_r)$; i.e. $\pstate_r=(\tau_r, \str^{*}(\tau_r))$.
We used the following sequence of intermediate distributions over forests:
$\gamma(\pstate_r) = \prod_{\tau_i\in \pstate_r}
\gamma(\tau_i, \str^{*}(\tau_i))).$
In the weight update step, besides proposing two branch lengths and randomly 
choosing a pair of trees from the current  forest to merge (as in \cite{Wang2012PhDThesis}), we   
use the proposal distribution described in the Methodology Section of the main paper for proposing the hidden strings, with the only difference that a root sequence is selected uniformly among the intermediate strings on the proposed path. Algorithm \ref{alg:TI} shows the component of the algorithm not present in previous work: proposing a root sequence and two discrete paths $L_\leftsub$ and $L_\rightsub$ linking this root to its two children. We assume that the pair of sequences to merge, $x_\leftsub$ and $x_\rightsub$, as well as the branch lengths connecting each to the newly formed root, $T_\leftsub, T_\rightsub$, have been picked by a standard phylogenetic SMC proposal, with proposal density $\tilde p_\treesub$. These become inputs to Algorithm~\ref{alg:TI}, which returns $(L_\leftsub, L_\rightsub, \tilde p, p)$. From this output, we compute two auxiliary rate matrices, $\check{Q}_\leftsub = \check{Q}(L_\leftsub)$ and $\check{Q}_\rightsub = \check{Q}(L_\rightsub)$, one for each newly formed branch, using the same method as before.

Given all this information, we update the particle weights as follows:\footnote{Note that this formula assumes that the process is reversible, but can be extended to the non-reversible case.}
\begin{align*} 
w_{g,k} = w_{g-1,k} \frac{p\ \textrm{statio}(x_\leftsub)}{\tilde p\ \tilde p_\treesub}\ (\exp(T_\leftsub \check{Q}_\leftsub))_{1,|L_\leftsub|}\ (\exp(T_\rightsub \check{Q}_\rightsub))_{1,|L_\rightsub|}.
\end{align*}

\begin{algorithm}[!htbp]
\caption{: {\bf phylo-sequence-proposal}($x_\leftsub, T_\leftsub, x_\rightsub, T_\rightsub$)}
\begin{algorithmic}
\STATE $(L, \tilde p, p) \gets$~{\bf propose}($x_\leftsub, x_\rightsub, T_\leftsub + T_\rightsub$)
\STATE $i \sim$~Unif$\{1, 2, \dots, |L|\}$
\STATE $\tilde p \gets \tilde p \times 1/|L|$
\STATE $L_\leftsub \gets$~{\bf reverse}({\bf sublist}($L, 1, i$))
\STATE $L_\rightsub \gets$~{\bf sublist}($L, i, |L|$)
\RETURN $(L_\leftsub, L_\rightsub, \tilde p, p)$
\end{algorithmic}
\label{alg:TI}
\end{algorithm}

\begin{figure}[!htbp]
\begin{center}
\includegraphics[width=0.6\textwidth]{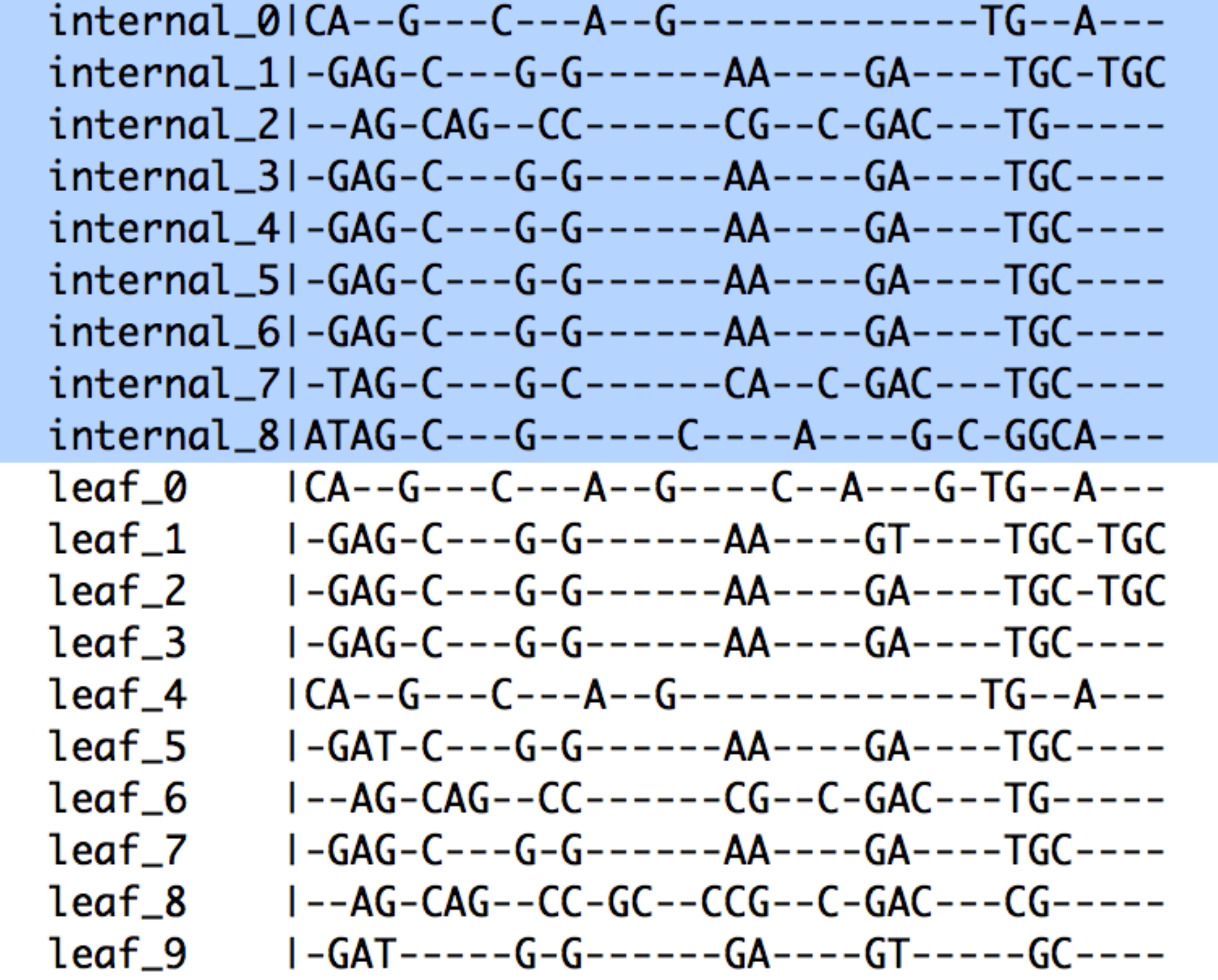}
  \caption{{\bf Sequence Simulation.} One subset of simulated sequence data for the experiments on trees via SMC.  
}
  \label{fig:simulatedData}
  \end{center}
\end{figure}

We simulated subsets of molecular sequences with
different random seeds according to our evolutionary model. The parameters 
are:  SSM length=3, 
$\theta_{sub}=0.03$, $\lambda_{pt}=0.05$,  $\mu_{pt}=0.2$, 
 $\lambda_{SSM}=2.0$, and 
 $\mu_{SSM}=2.0$.  A subset of the data is shown in the Figure~\ref{fig:simulatedData}.
 
\subsection{RNA Folding Pathways}   To compare our method (denoted as TIPS) to that of forward sampling (denoted as FS), we first obtained an absolute accuracy by computing the matrix exponential. Then we computed the absolute log error estimate $\hat p$ (i.e., error($\hat p$)~$=|\log \hat p - \log \P_x(X_N = y)|$) of our method and forward sampling on the RNA molecules shown in Table~\ref{tab:RNAseq}.  These RNA sequences are short, due to the limit in the size of matrix computed using the matrix exponential.  The state spaces for the first two RNA sequences were sufficiently small for computation of the matrix exponential.  The complete state space of the last two RNAs were too large for the matrix exponential, so we sampled a subset of the state space.

\begin{table}[!hptb]
\begin{center}
\large
\begin{tabular}{ l | l  l  l}
  Sequence &  Length & $|\states|$ & $|S|$ \\
  \hline                        
  1AFX & 12 & 70 & - \\    
  1XV6 & 12 & 48 & - \\     
  RNA21 & 21 & $\sim{1100}$ & 657\\   
	HIV & 23 & $\sim{1500}$ & 266\\
  \hline  
\end{tabular}
\caption{\bf Biological RNA Sequences. }
\label{tab:RNAseq}
\end{center}
\end{table}

\begin{figure*}[!htbp]
\begin{center}
	\subfloat[\tiny{1AFX- error}]{\label{fig: 1AFX-s}\includegraphics[width=.45\textwidth,height=0.147\textheight]{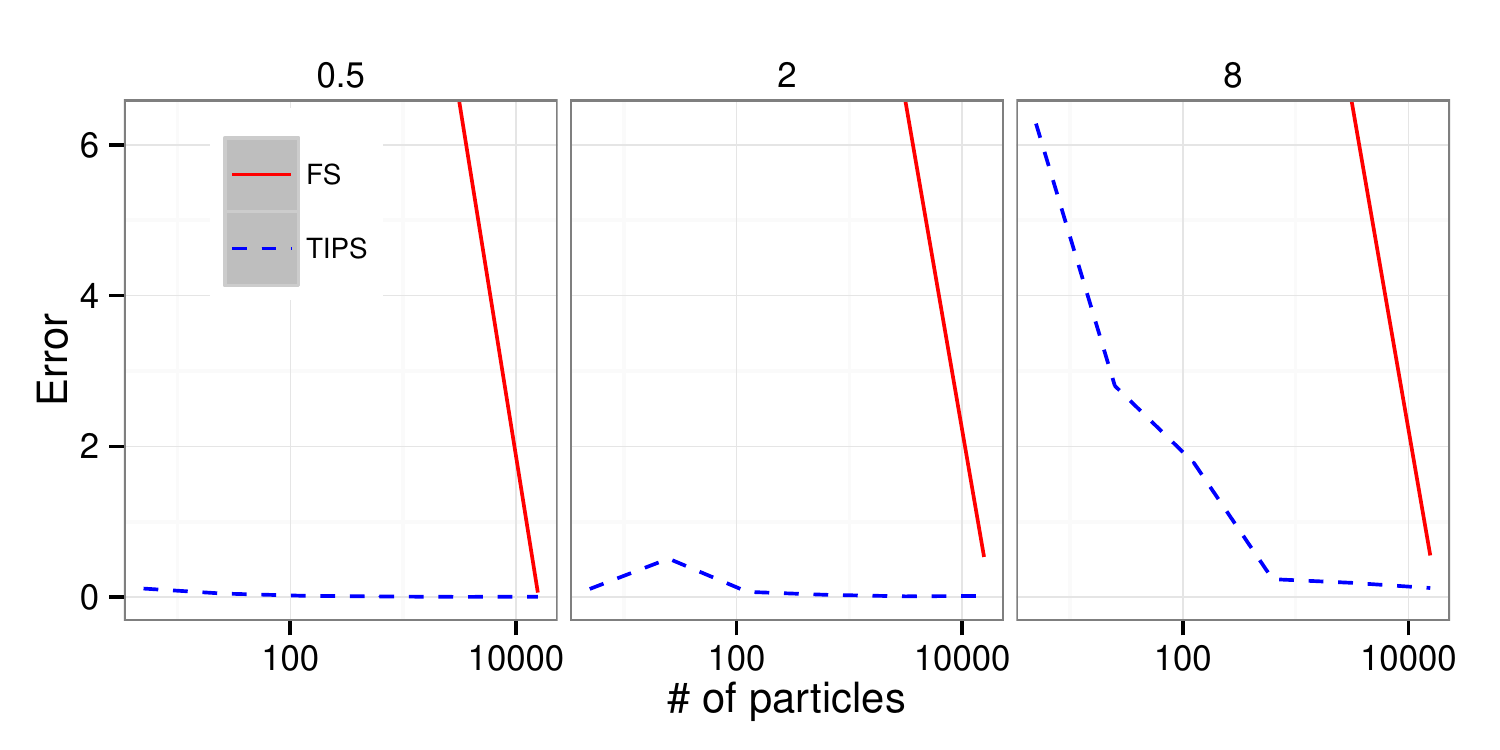} }
		\subfloat[\tiny{1AFX-~CPU time~(ms)}]{\label{fig: 1AFX-r}\includegraphics[width=.2\textwidth,height=0.14\textheight]{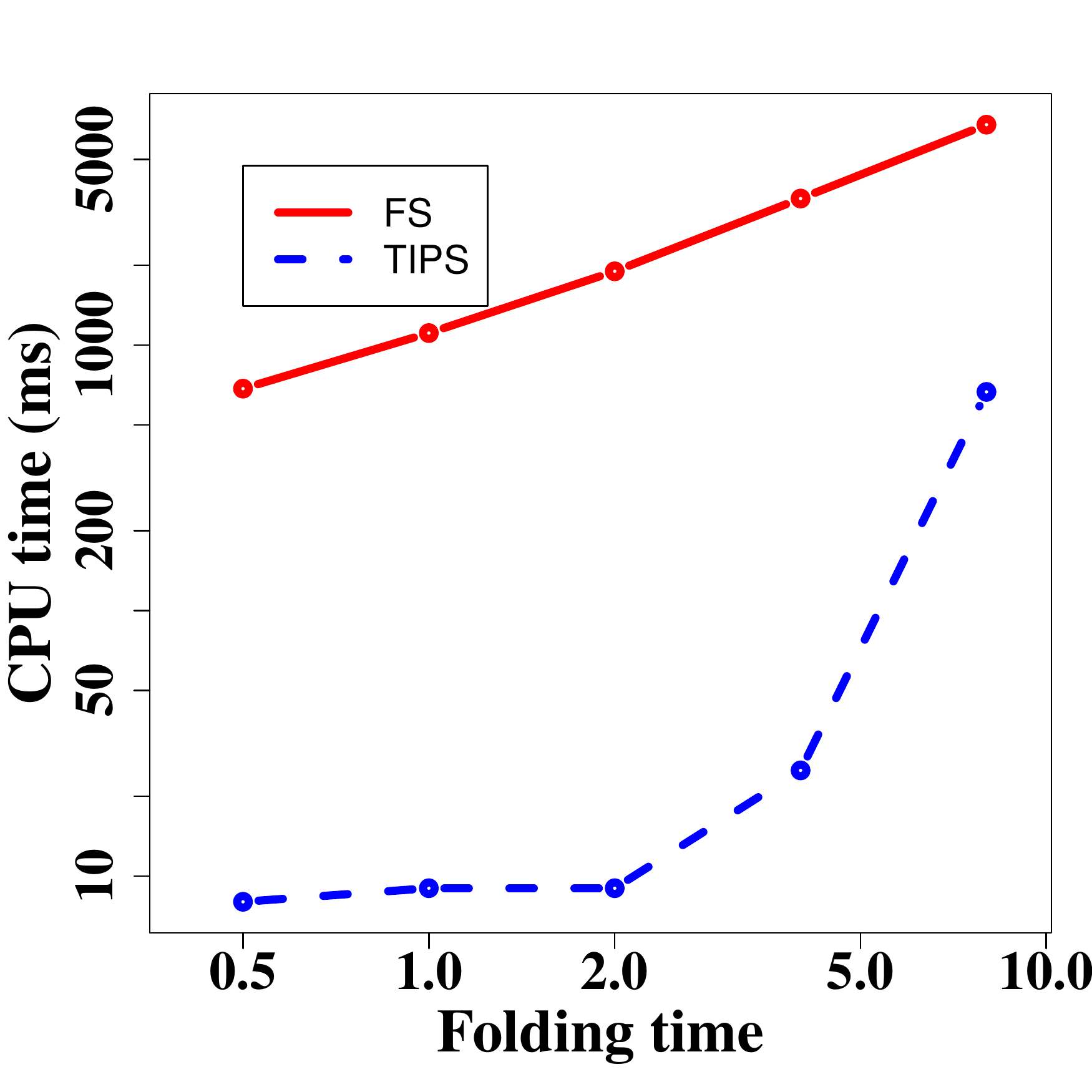} }
		\subfloat[\tiny{1AFX- variance}]{\label{fig: 1AFX-s}\includegraphics[width=.2\textwidth,height=0.14\textheight]{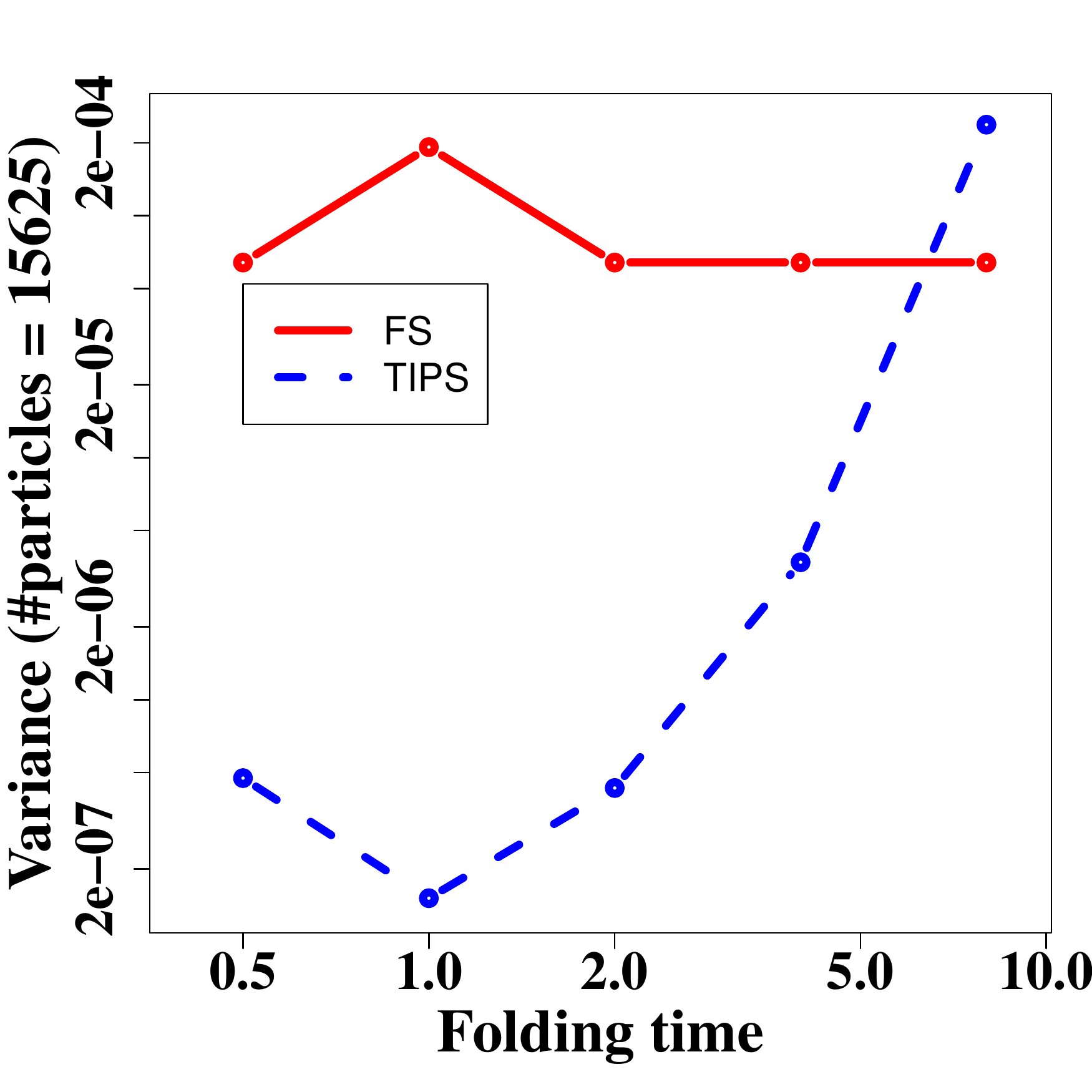}}
		
\subfloat[\tiny{1XV6~- error}]{\label{fig: 1XV6-s}\includegraphics[width=.45\textwidth,height=0.147\textheight]{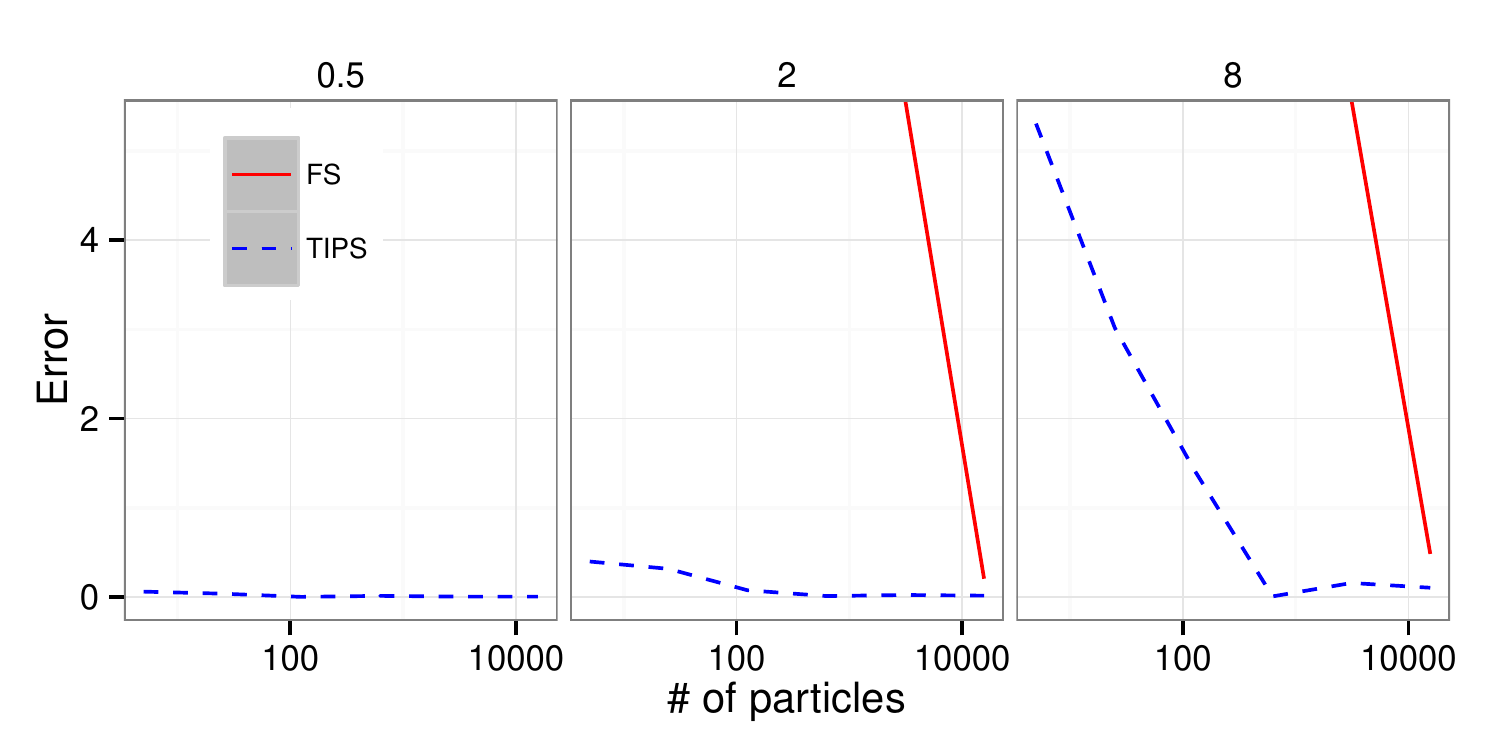} }
	\subfloat[\tiny{1XV6~- CPU time~(ms)}]{\label{fig: 1XV6-r}\includegraphics[width=.2\textwidth,height=0.14\textheight]{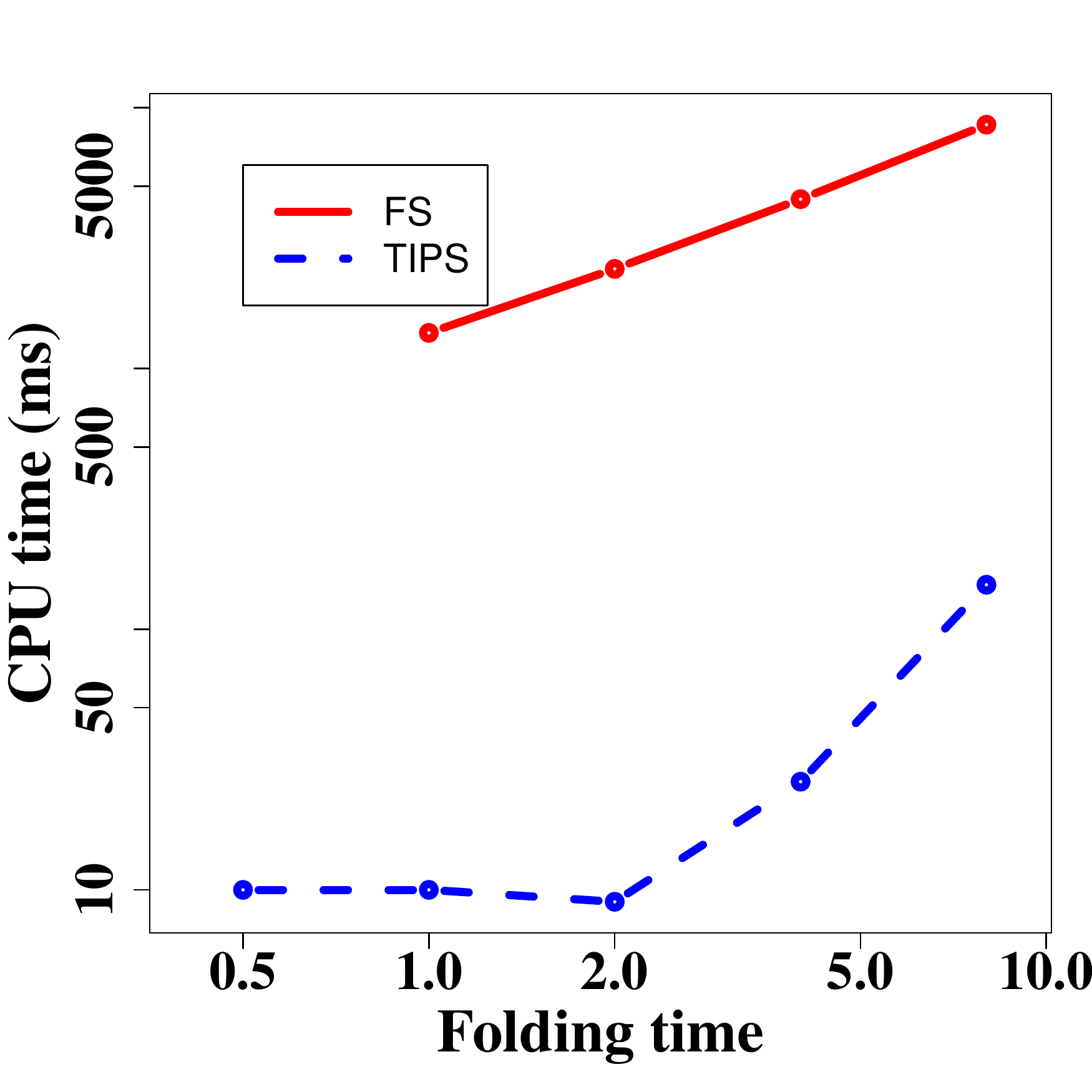} }
	\subfloat[\tiny{1XV6~- variance}]{\label{fig 1XV6-s}\includegraphics[width=.2\textwidth,height=0.14\textheight]{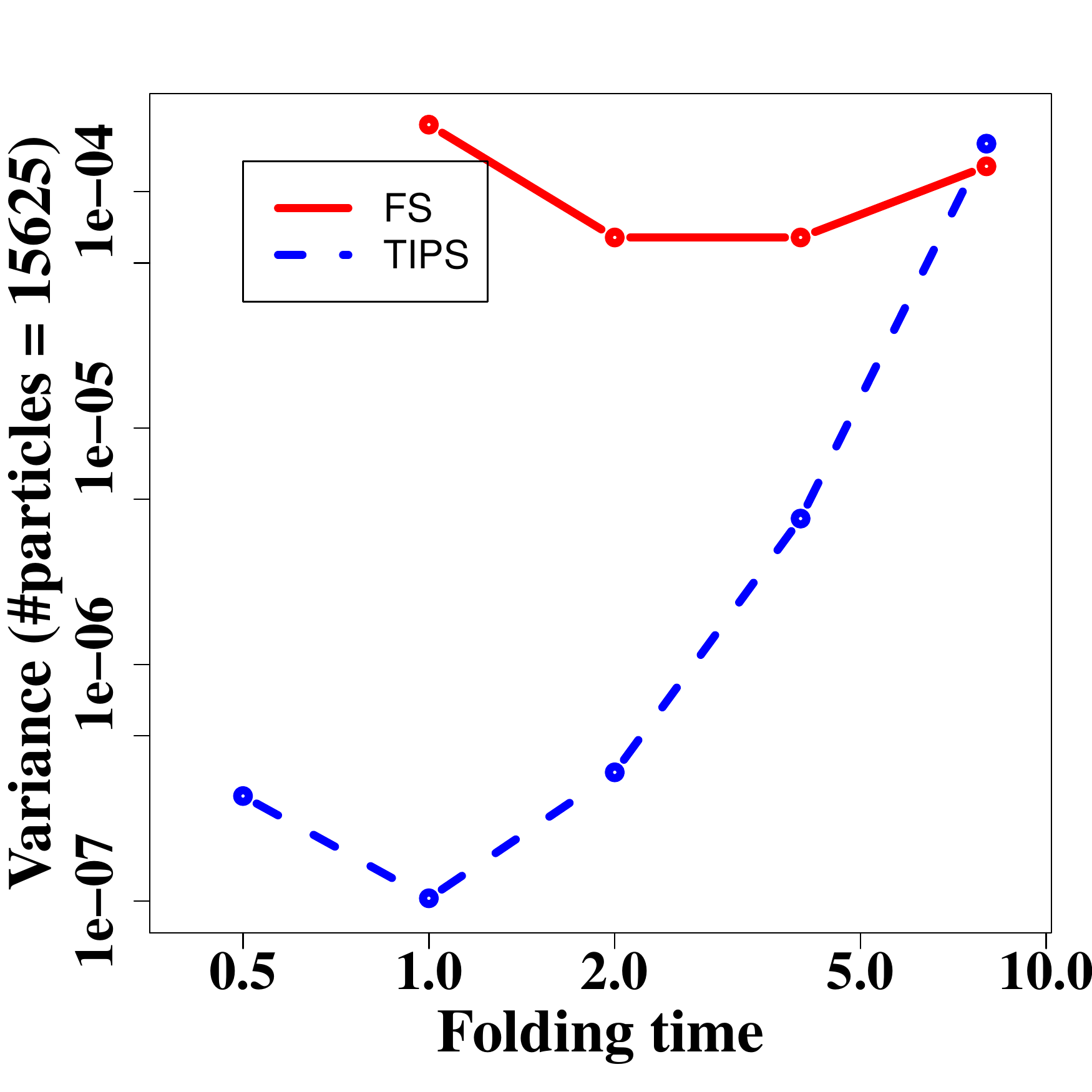}}

\caption[Performance vs. folding time]{Performance of our method (TIPS) and forward sampling (FS) on 1AFX and 1XV6 molecules with their \emph{whole} state space. The relative errors of the estimates vs. folding times, \{0.5,2,8\} are shown (left) along with the CPU times corresponding to the minimum number of particles required to satisfy the accuracy level $I$ in milliseconds (middle) and the variance of TIPS and FS estimations (right) on folding times, $\{0.5, 1,\cdots, 8\}$.}
\label{fig:RNAseq2}
\end{center}
\end{figure*}

Figures \ref{fig:RNAseq2}a, \ref{fig:RNAseq2}d  show the performance of the FS and TIPS methods, for two short molecules 1AFX and 1XV6, on selective folding times, $\{0.5,2,8\}$. Figures \ref{fig:RNAseq2}b, \ref{fig:RNAseq2}e show the CPU times (in milliseconds) corresponding to the minimum number of particles required to satisfy the certain accuracy level, $I=$~\{$\hat p:$ error($\hat p$)~$< 1.0$\} on different folding times including the selective ones. 

The variances of FS and TIPS weights, for $5^6 = 15625$ particles, are also computed and compared on different folding times (see Figures \ref{fig:RNAseq2}c, \ref{fig:RNAseq2}f).

\begin{figure}[!htbp]
\begin{center}
\subfloat{\label{fig: 1XV6-s-t}\includegraphics[width=.5\textwidth]{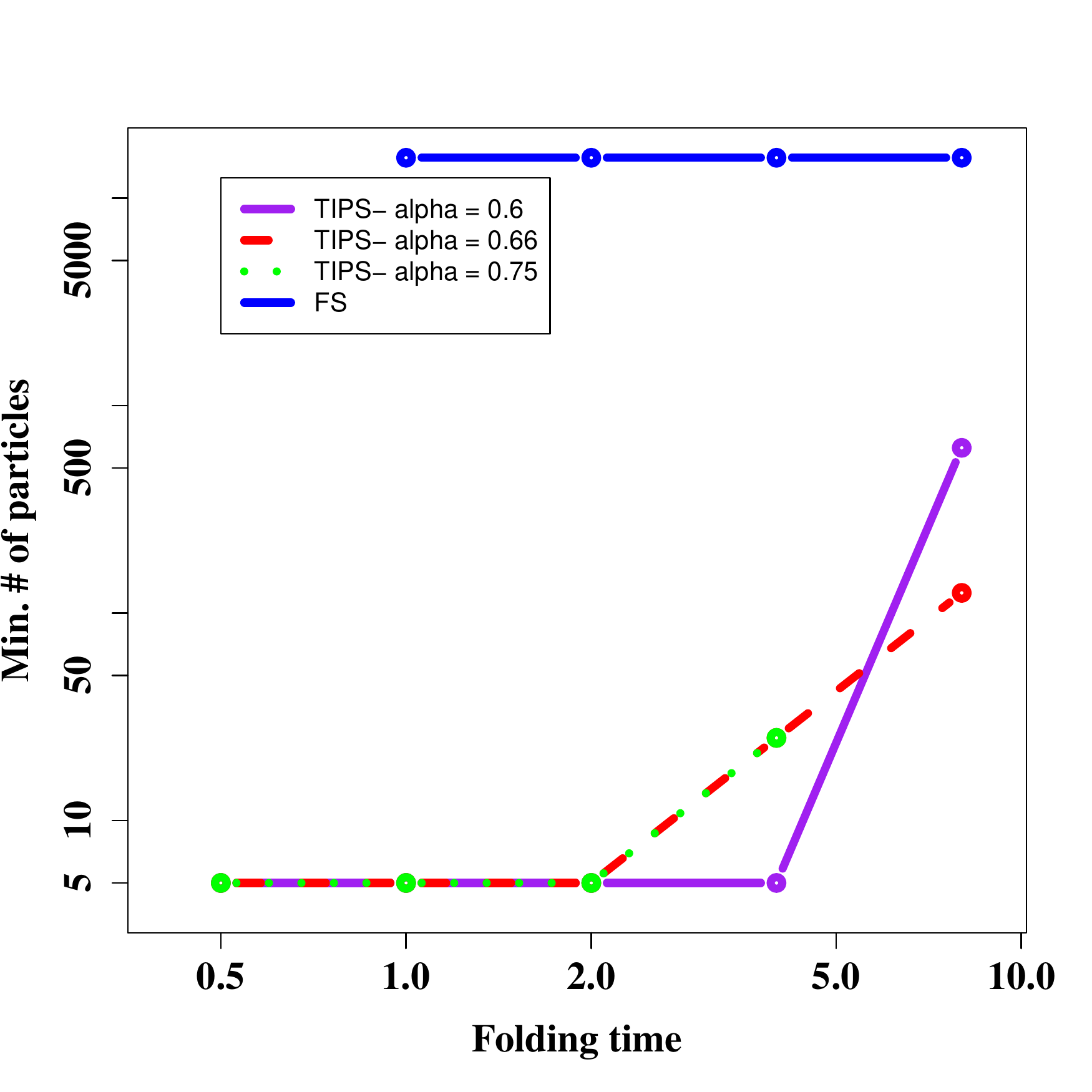}}

\caption{{\bf Tuning parameter $\alpha$. }Performance of our method (TIPS) using different values of $\alpha$ compared to forward sampling (FS) for estimating the folding pathway of the 1XV6 molecule on its \emph{whole} state space.}
\label{fig:RNAseq2-tuning}
\end{center}
\end{figure}

Our method has two main tuning parameters, a geometric parameter, $\beta$, over the number of repeated excursions from $y$ to itself, and a parameter, $\alpha$, weighting the probability of the steps decreasing the value of the potential.  We found that the accuracy of the sampler in the case of RNA folding pathways was sensitive to the setting of the parameters (see Figure~\ref{fig:RNAseq2-tuning}).  Parameter tuning is an important area of future work.

\begin{figure*}[!t]
\begin{center}
\subfloat{\includegraphics[width=.18\textwidth, height = .07\textheight]{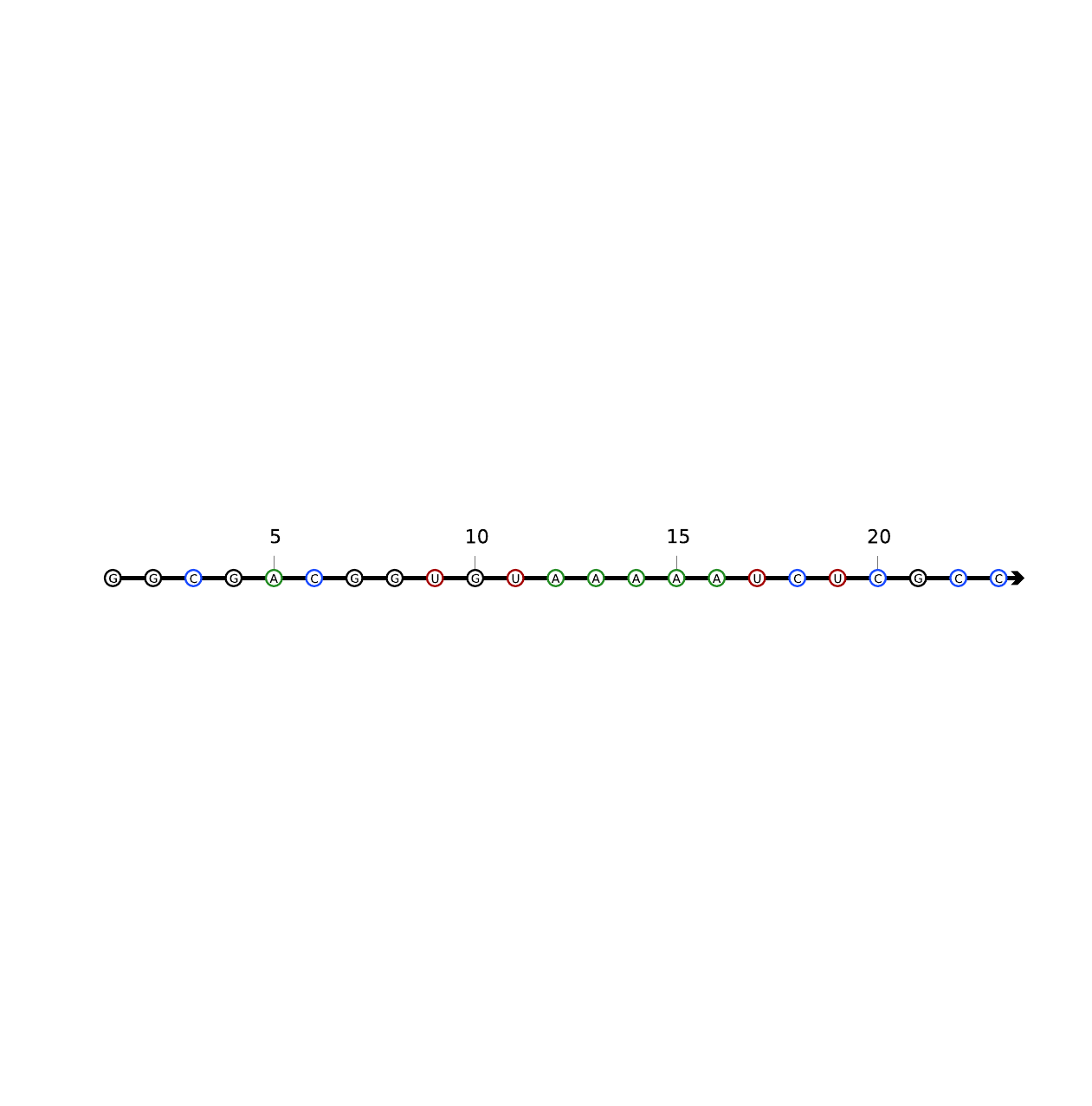}}
\subfloat{\includegraphics[width=.11\textwidth, height = .07\textheight]{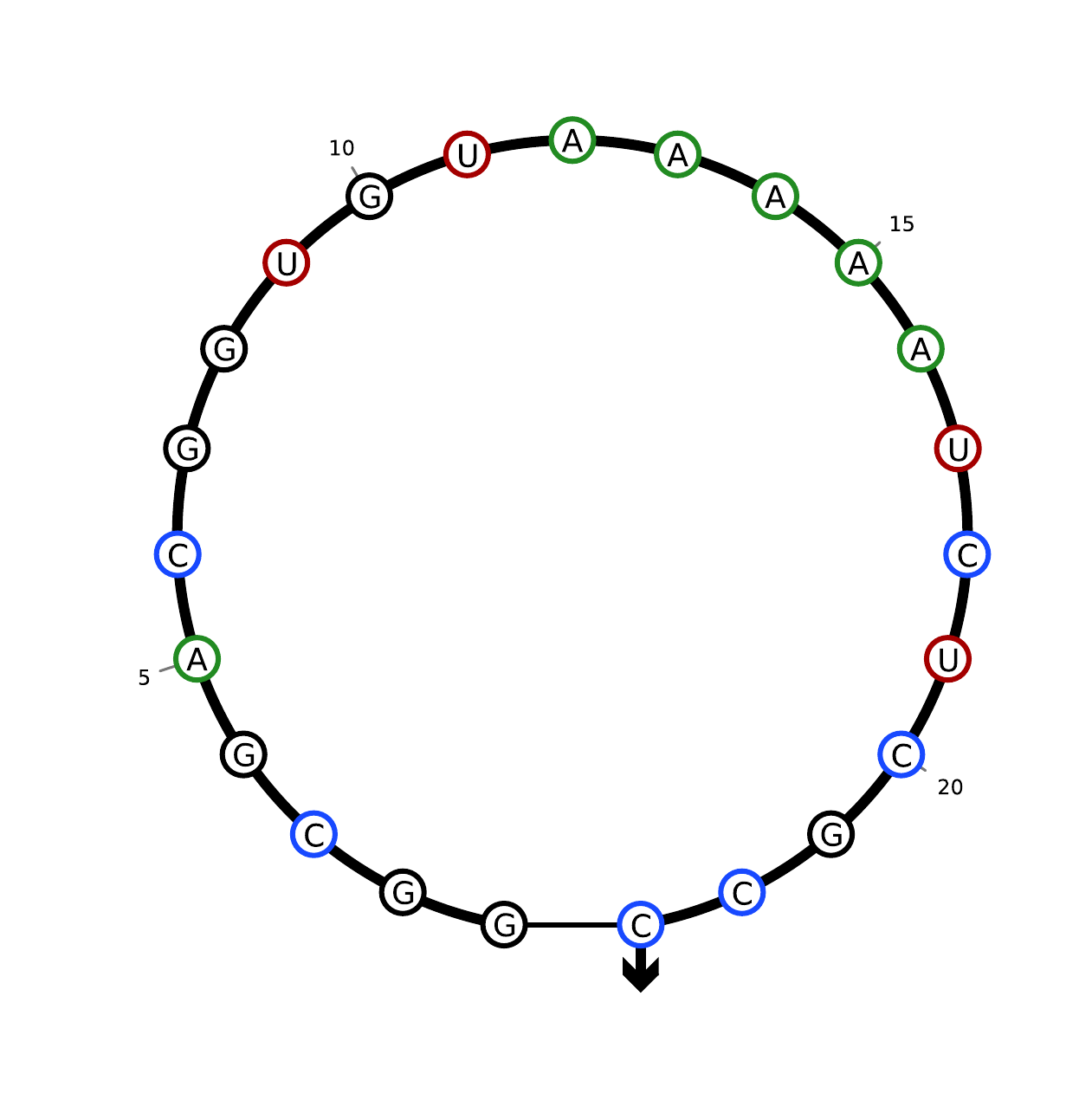}}
\subfloat{\includegraphics[width=.11\textwidth, height = .07\textheight]{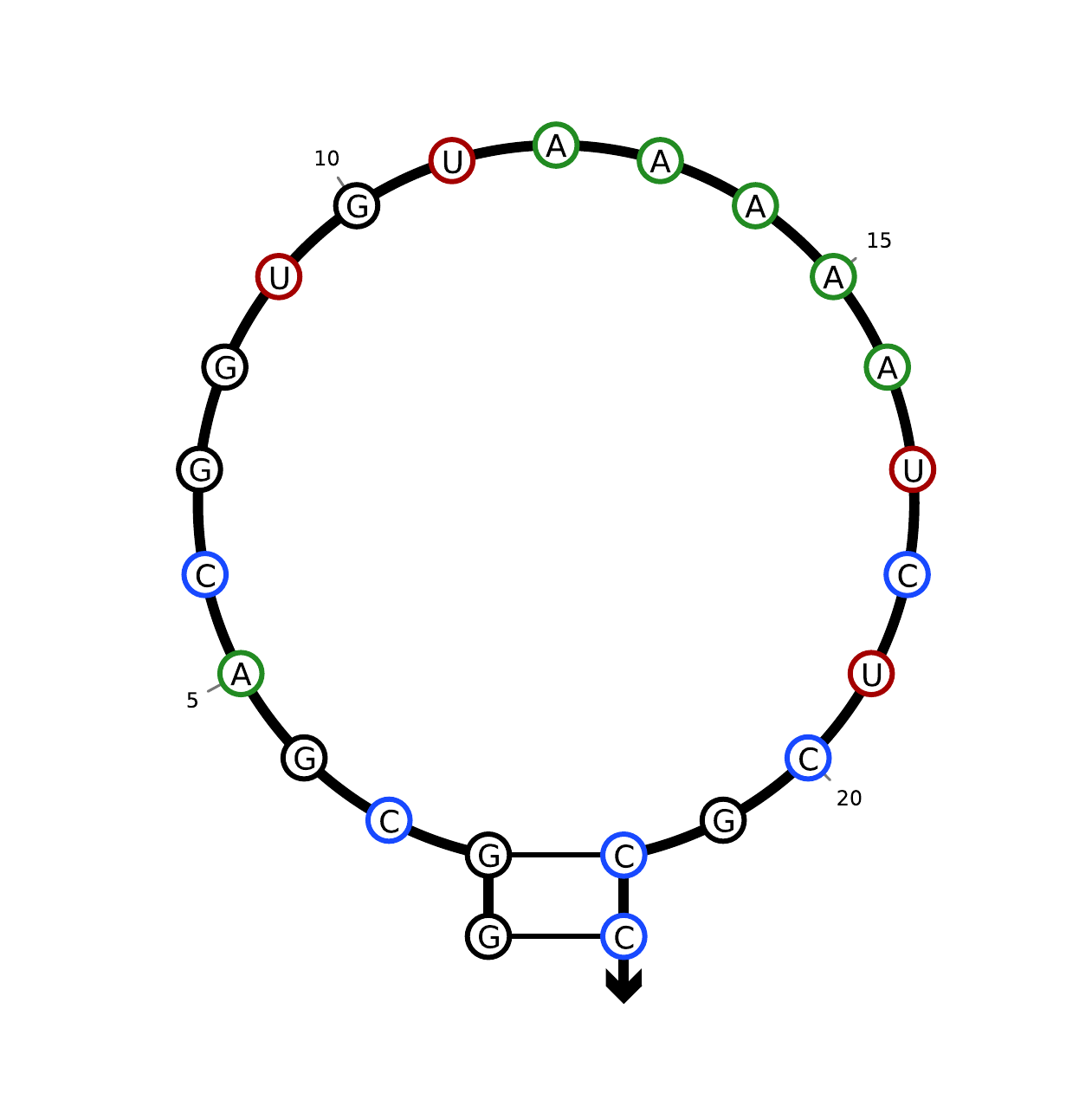}}
\subfloat{\includegraphics[width=.11\textwidth, height = .07\textheight] {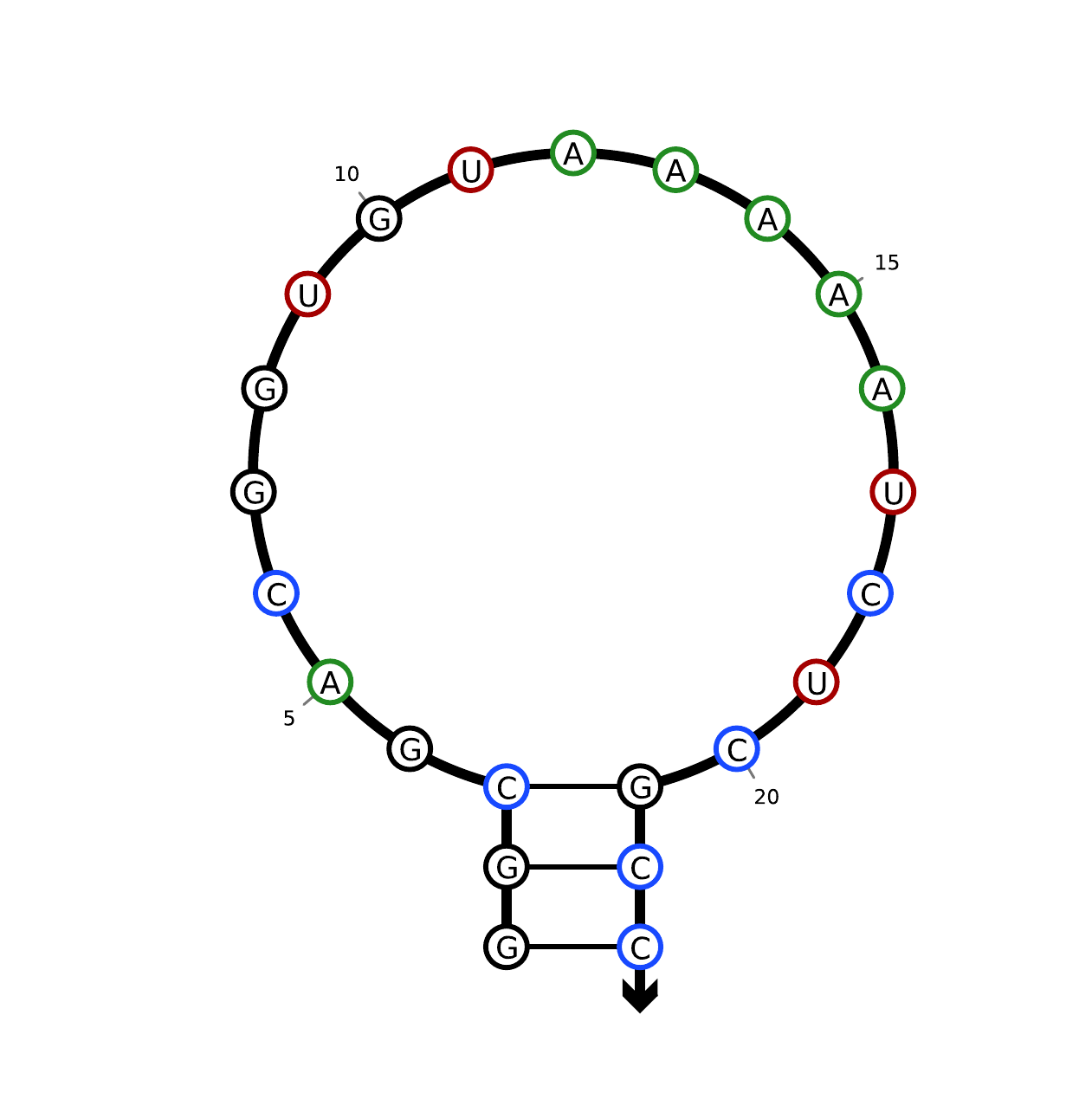}}
\subfloat{\includegraphics[width=.11\textwidth, height = .07\textheight]{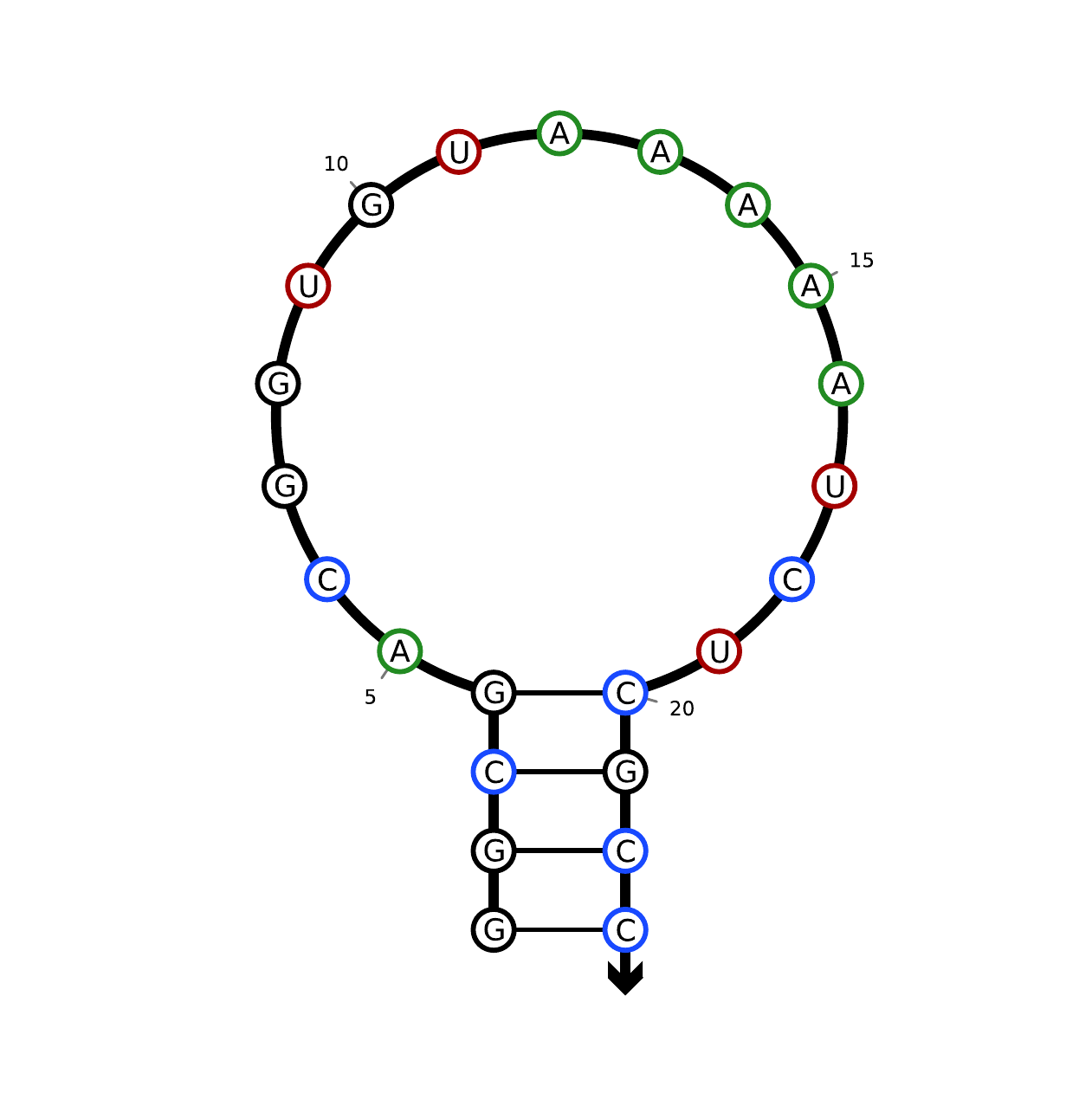}}
\subfloat{\includegraphics[width=.11\textwidth, height = .07\textheight]{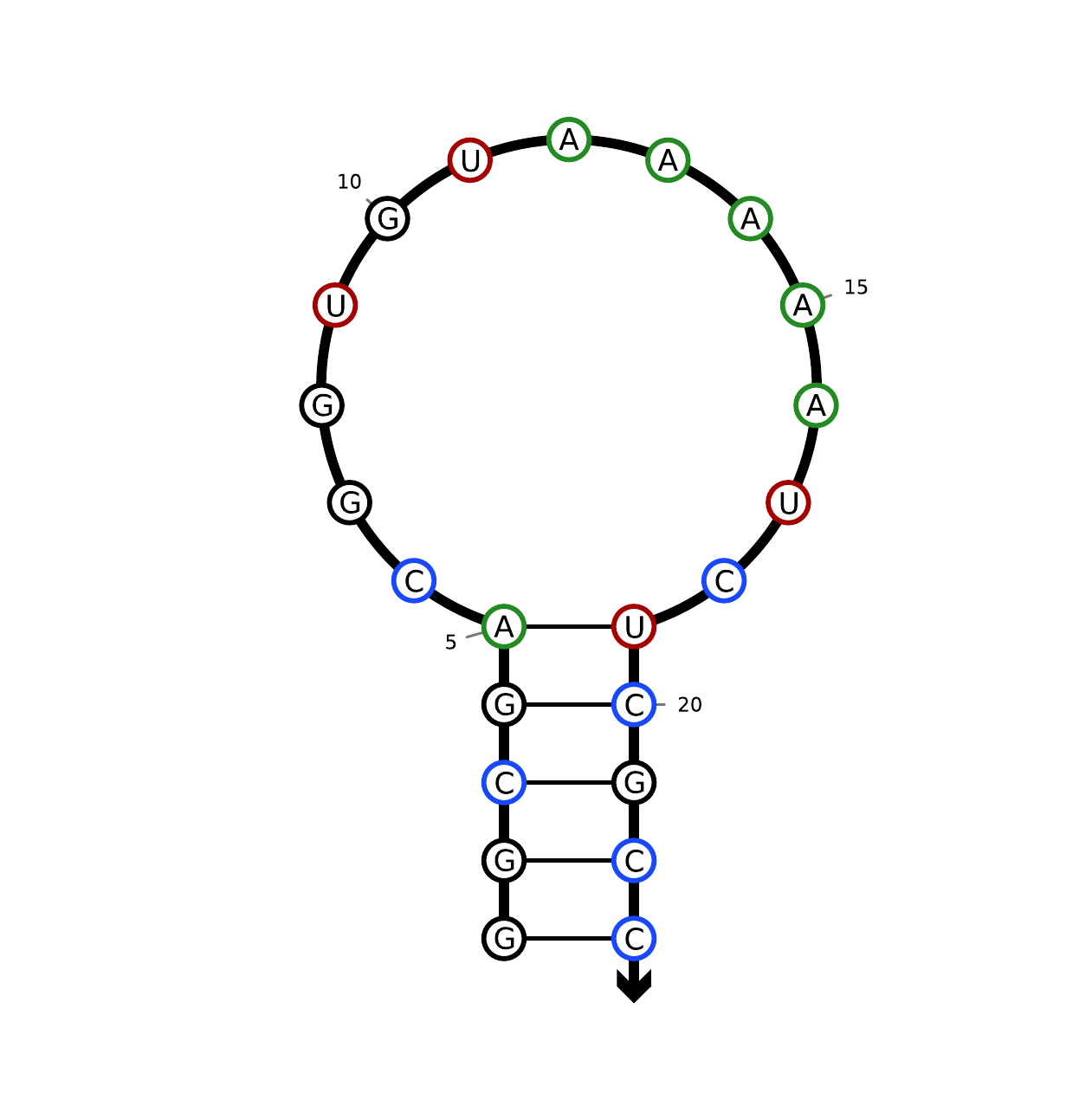}}
\subfloat{\includegraphics[width=.11\textwidth, height = .07\textheight]{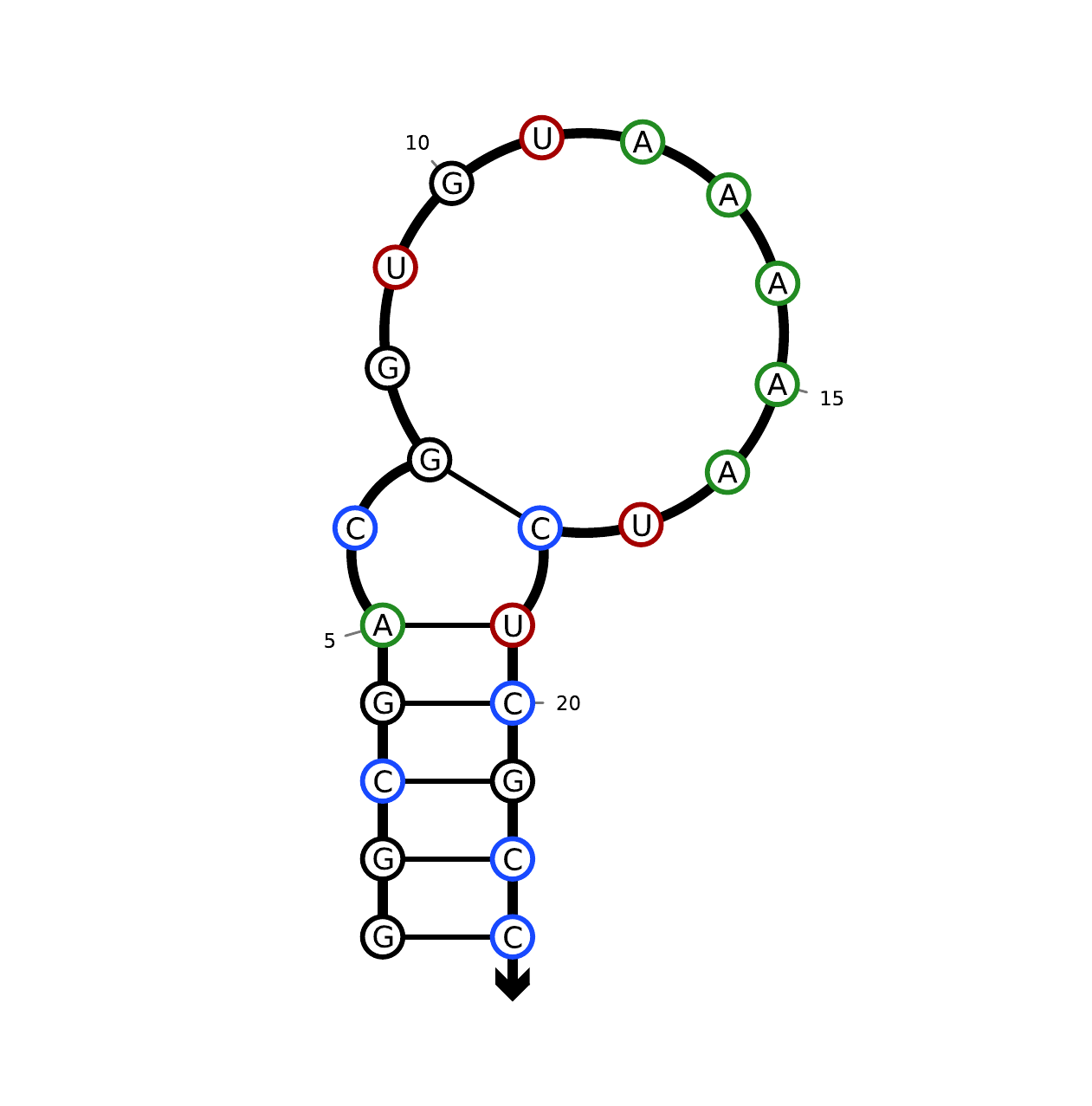}}
\subfloat{\includegraphics[width=.11\textwidth, height = .07\textheight]{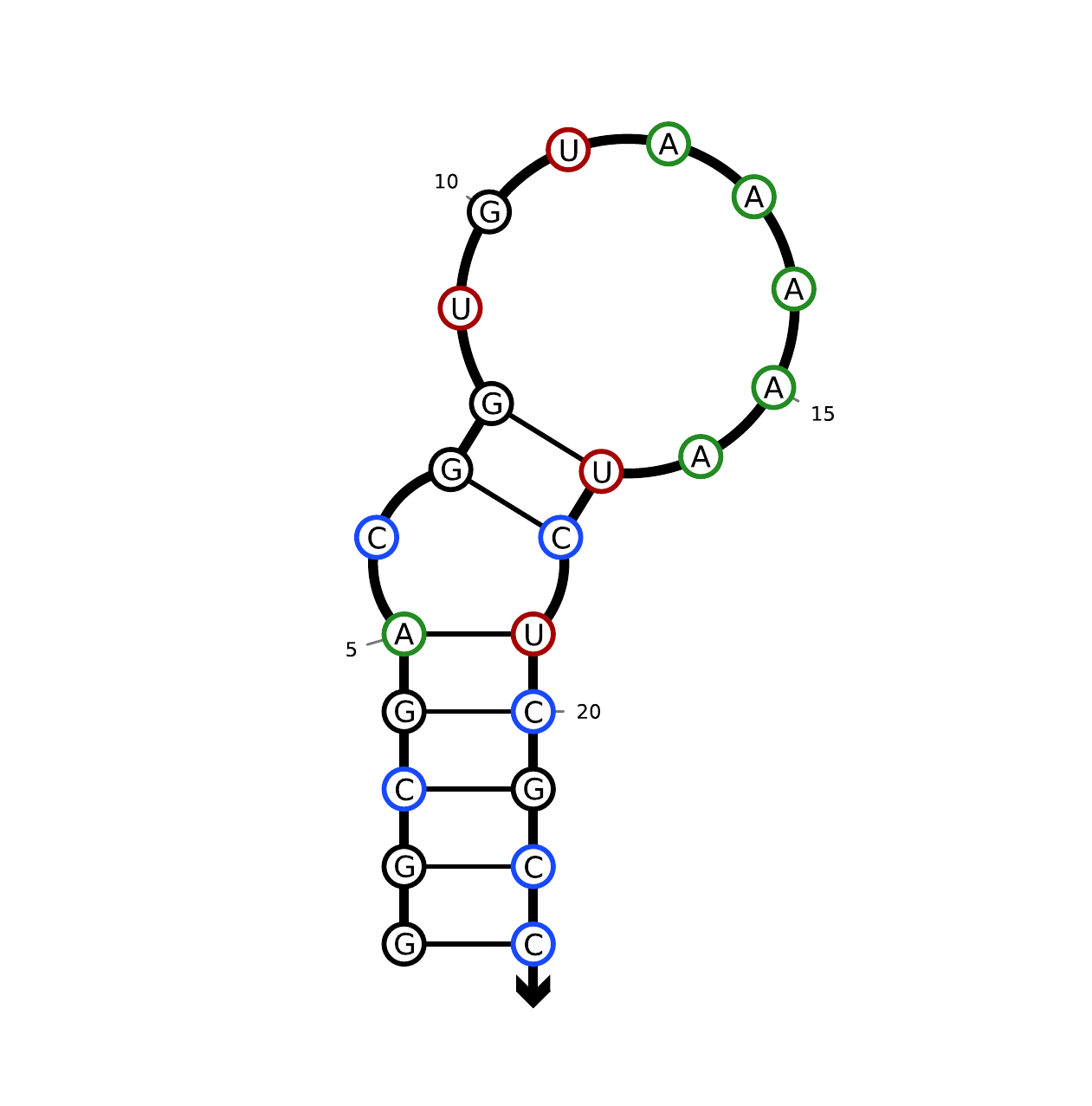}}
\subfloat{\includegraphics[width=.11\textwidth, height = .07\textheight]{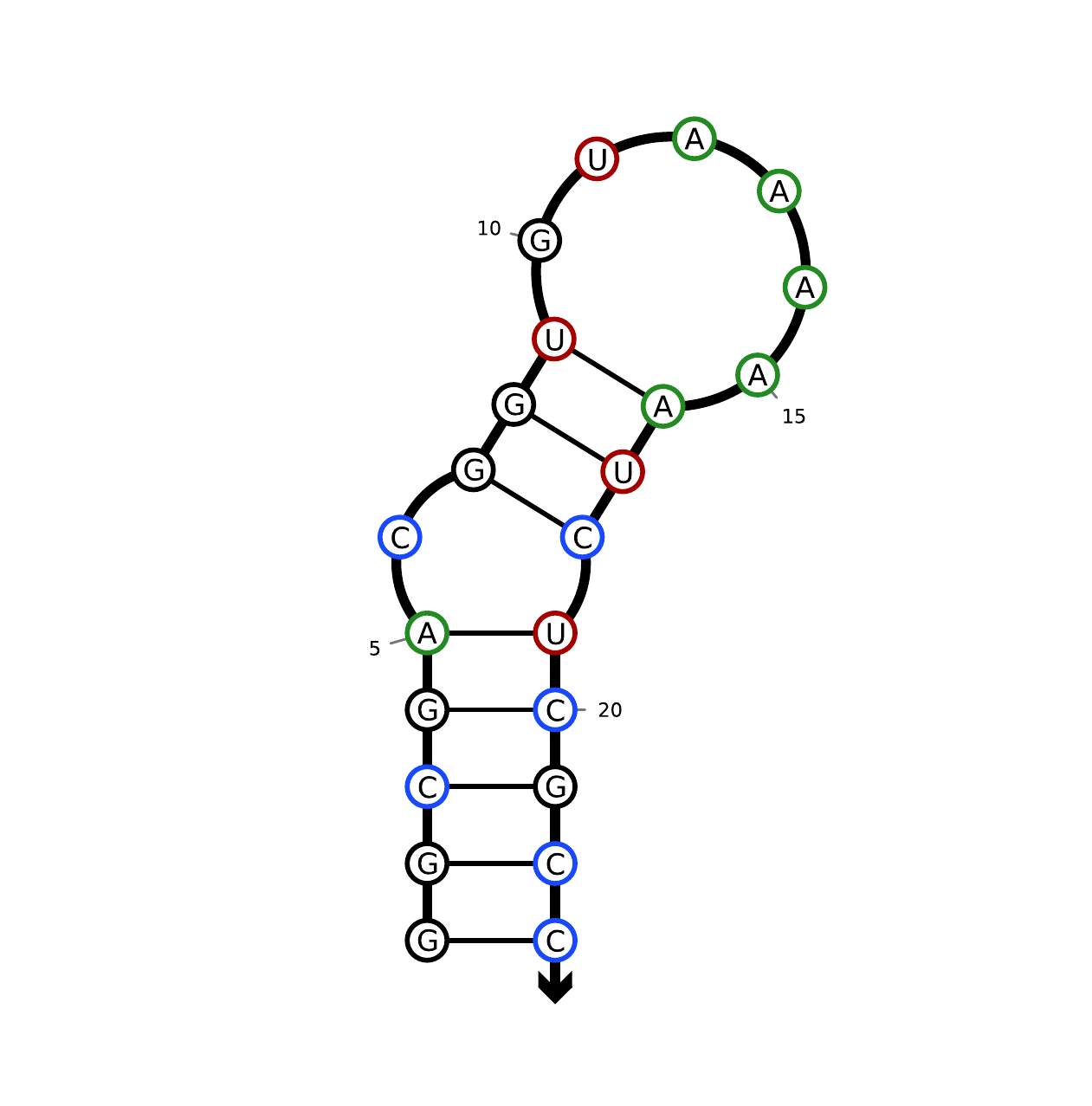}}
\caption{An example of a sampled folding pathway for the HIV23 molecule with $T=0.125$.}
\label{fig:HIV23_IS_path}
\end{center}
\end{figure*}



\nocite{Akkouchi2008}
\nocite{Amari1997SumExp}
\nocite{Andrieu2009PseudoMarg}
\nocite{Andrieu2010}
\nocite{Arribas-Gil2012}
\nocite{Beaumont2003GIMH}
\nocite{Bouchard2012Evolutionary}
\nocite{Bouchard2012Phylogenetic}
\nocite{fan2008sampling}
\nocite{Felsenstein1981}
\nocite{Felsenstein2003}
\nocite{Flamm2000}
\nocite{Grassmann1977Uniformization}
\nocite{Hickey2011b}
\nocite{huelsenbeck_mrbayes:_2001}
\nocite{Hutter2009}
\nocite{Juneja2006RareEvent}
\nocite{kirkpatrick_new_2013}
\nocite{Lakner01022008}
\nocite{Morrison2009}
\nocite{Munsky2006}
\nocite{Rao2011}
\nocite{Rao2012}
\nocite{Saeedi2011Priors}
\nocite{Schaeffer2012Multistrand}
\nocite{teh08a}
\nocite{Venkataraman2010}
\nocite{Wang2012PhDThesis}
\nocite{Wang2013AHMC}
\newpage
{\footnotesize
\bibliography{bib}
\bibliographystyle{icml2014}
}